\documentclass[aoas]{imsart}

\usepackage{lscape, rotating}   
\usepackage{booktabs}   
\usepackage{textgreek}
\usepackage{ccaption}
\usepackage{ragged2e} 
\usepackage{mathtools, stackrel, enumerate}
\usepackage{tabularx}
\usepackage[stable]{footmisc}
\usepackage{floatrow}
\RequirePackage{epigraph}
\RequirePackage{amsthm,amsmath,amsfonts,amssymb}
\RequirePackage[numbers]{natbib}
\RequirePackage[colorlinks,citecolor=blue,urlcolor=blue]{hyperref}
\RequirePackage{graphicx, subcaption}

\startlocaldefs
\theoremstyle{plain}

\theoremstyle{remark}

\def\bx{\boldsymbol{x}}
\def\bX{\boldsymbol{X}}
\def\btheta{\boldsymbol{\theta}}

\endlocaldefs

\setcitestyle{authoryear,round}
\begin{document}

\begin{frontmatter}
\title{The Roles and Challenges of the \textit{P}-value\thanksref{T1}}

\runtitle{On \textit{P}-value}

\thankstext{T1}{An early version of this paper was posted in arXiv in 2020 (arXiv:2002.07270) entitled \textit{Thou Shalt Not Reject the \textit{P}-value}.}

\begin{aug}
\author[A]{\fnms{Oliver Y.} \snm{Ch\'en}\ead[label=e1]{olivery.chen@bristol.edu}}, 
\author[B]{\fnms{Ra\'ul G.} \snm{Saraiva}\ead[label=e2]{}},
\author[C]{\fnms{Huy} \snm{Phan}\ead[label=e3]{}},
\author[D]{\fnms{Junrui} \snm{Di}\ead[label=e4]{}},
\author[E]{\fnms{Guy} \snm{Nagels}\ead[label=e5]{}},
\author[F]{\fnms{Tom} \snm{Schwantje}\ead[label=e6]{}},
\author[G]{\fnms{Hengyi} \snm{Cao}\ead[label=e7]{}},
\author[H]{\fnms{Jiangtao} \snm{Gou}\ead[label=e8]{}},
\author[I]{\fnms{Jenna M.} \snm{Reinen}\ead[label=e9]{}},
\author[J]{\fnms{Bin} \snm{Xiong}\ead[label=e10]{}},
\author[K]{\fnms{Bangdong} \snm{Zhi}\ead[label=e10]{}},
\author[K]{\fnms{Xiaojun} \snm{Wang}\ead[label=e10]{}}
\and
\author[L]{\fnms{Maarten} \snm{de Vos}\ead[label=e11]{}\ead[label=u1,url]{}}
\address[A]{Faculty of Social Sciences and Law, University of Bristol, \printead{e1}}

\address[B]{Apple Tree Partners}

\address[C]{Department of Computer Science, Queen Mary University of London}

\address[D]{Pfizer}

\address[E]{Department of Engineering, University of Oxford}

\address[F]{Department of Economics, University of Oxford}

\address[G]{Department of Psychology, Yale University}

\address[H]{Department of Mathematics and Statistics, Villanova University}

\address[I]{IBM Thomas J. Watson Research Center}

\address[J]{Department of Statistics, Northwestern University}

\address[K]{School of Management, University of Bristol}

\address[L]{Faculties of Engineering Science and Medicine, KU Leuven}

\end{aug}

\begin{abstract}
Since its debut in the 18th century, the \textit{P}-value has been an important part of hypothesis testing-based scientific discoveries. As the statistical engine accelerates, questions are beginning to be raised, asking to what extent scientific discoveries based on \textit{P}-values are reliable and reproducible, and the voice calling for adjusting the significance level or banning the \textit{P}-value has been increasingly heard. Inspired by these questions and discussions, here we enquire into the useful roles and misuses of the \textit{P}-value. For common misuses and misinterpretations, we provide modest recommendations. In parallel, we present the Bayesian alternatives for seeking evidence. Finally, we discuss the promises and risks of using meta-analysis to pool \textit{P}-values from multiple studies to aggregate evidence. Taken together, the \textit{P}-value underpins a useful probabilistic decision-making system and provides evidence at a continuous scale. But its interpretation must be contextual, considering the scientific question, experimental design (including the model specification, sample size, and significance level), statistical power, effect size, and reproducibility.
\end{abstract}



\end{frontmatter}

\epigraph{
Most statisticians are all too familiar with conversations [that] start [with]:\\
\textit{Q}: \textit{What is the purpose of your analysis?}\\
\textit{A}: \textit{I want to do a significance test.} \\
\textit{Q}: \textit{No, I mean what is the overall objective?} \\
\textit{A} (with puzzled look): \textit{I want to know if my results are significant.} \\
And so on ...\\
\vspace{2mm}
\citep{angrist1996identification}
}

\section{Introduction}

David Hume argued in \textit{A Treatise of Human Nature} that “all knowledge degenerates into probability” \citep{hume1738treatise}. In humans, the probable inference is chief in guiding decisions \citep{nagel1939probability, morgenstern1944theory, tversky1994support}. Sports fans make bets on the likelihood that a club will win the next game. Investors decide to buy or sell a stock based on how likely it is to go up or down. One chooses whether to bring an umbrella given the chance of rain. 
But what about scientists? How does probability guide scientific enquiries \citep{de1989probabilism}? 

A widely
\footnote{Text mining using 385,393 PubMed Central (PMC) articles from 1990 to 2015 identified 3,438,299 appearances of \textit{P}-values; that is, about nine \textit{P}-values per article \citep{chavalarias2016evolution}.} 
used principle in scientific decision-making is the \textit{P}-value-based hypothesis testing. It has interested biomedical scientists \citep{panagiotakos2008value}, clinicians \citep{singh2008interpreting}, social scientists \citep{skipper1967sacredness}, and philosophers \citep{richard2017statistical}, no less than statisticians. 

Yet, as a probabilistic statement underpinning decision-making, the \textit{P}-value has generated enduring debates \citep{berger1987testing, casella1987reconciling, cohen1994earth, greco2011significance, harlow2013if, ziliak2008cult, spanos2010frequentist}. Central to these debates are its inconsistency and potential lack of credibility for providing evidence. To raise protection, scholars have suggested lowering the significance level from $0.05$ to $0.005$ \citep{benjamin2018redefine, ioannidis2018proposal}. Others have asked whether the \textit{P}-value (and the significance test) should be banned \citep{hunter1997needed, kraemer2019time, shrout1997should}. The \textit{Basic and Applied Social Psychology} journal, at perhaps the extreme end, cast an editorial ban on the \textit{P}-value \citep{trafimow2015editorial}.

These debates and responses inspired us to have a thorough reflection on and discussion about the \textit{P}-value, from its origin and definition to its usefulness, misuses, and potential mitigations. We are fortunate to have access to many past works on the \textit{P}-value from statistics, biology, medicine, and philosophy \citep{de1989probabilism, gelman2011induction,leek2015statistics, tversky1994support}. Standing on their shoulders, we make our addition. 

We begin with a brief history of the \textit{P}-value. We then outline the roles the \textit{P}-value plays in scientific enquires, and, particularly, in causal inference. Next, we present its common misuses and misinterpretations with a discussion on potential treatments. Subsequently, we compare statistical significance with clinical relevance. In parallel, we examine the Bayesian alternatives for seeking evidence and discuss the promises and dangers of using meta-analysis to pool \textit{P}-values from multiple studies and datasets. We conclude with a discussion, and stress that one needs to employ and interpret the \textit{P}-value in context, considering the scientific question, experimental design (including the model specification, sample size and significance level), statistical power, effect size, and reproducibility.

\section{A brief history of the \textit{P}-value}

\subsection{The debut of the \textit{P}-value}
Although the origin of the hypothesis test and the \textit{P}-value is difficult to trace, John Arbuthnot performed the first known significant testing (p.40, \citep{heyde2001statisticians}). Having observed that the number of males born in London exceeded the number of females for $82$ consecutive years ($1629 - 1710$), Arbuthnot wanted to examine whether the birth rates of males and females were equal. He assumed two hypothetical individuals $A$ and $B$, where $B$ claimed that “… every year there shall be born more Males than Females” and $A$ laid a hypothesis against $B$’s. He then argued that if the birth rates were equal, the probability of observing more male new-borns for $82$ consecutive years would be $0.5^{82}$ \citep{arbuthnot1712ii}. Based on this infinitely small likelihood, he concluded that the birth rates were not equal. It was a relatively simple sign test, but a remarkable step in statistical history\footnote{Indeed, it was “the first example of reasoning about statistical significance” \citep{hald1998history} and “perhaps the first published report of a nonparametric test” \citep{conover1999practical}.}.

\subsection{The rise of the \textit{P}-value and hypothesis testing}

“Throughout the 19th century, hypothesis testing was carried out rather informally without a prespecified rejection level. It was roughly equivalent to calculating a (approximate) \textit{P}-value and rejecting the hypothesis if this value appeared to be sufficiently small” \citep{lehmann1993fisher}. Francis Edgeworth and Karl Pearson advanced the practice of significance tests during the late 19th and early 20th century. The former designed a test to compare means from two samples \citep{edgeworth1885observations} and introduced the concept of standard distance and a rejection rule (Chapter 13, \citep{spanos2019probability}). The latter introduced the \textit{Chi}-square test, and calculated the tail probability (which he denoted as capital \textit{P}) by integration \citep{pearson1900criterion}. Edgeworth implicitly used the tail probability, or the \textit{P}-value, in his test, followed by Pearson's formalization. It is, therefore, reasonable to credit them, in concert, the very considerable contribution in establishing the concept of the \textit{P}-value. 

The next milestone was made by R.A. Fisher in his seminal work \citep{fisher1925statistical}. He argued that “the [critical] value for which \textit{P} $= .05$, or $1$ in $20$, is $1.96$ or nearly $2$ [standard deviations]; it is convenient to take this point as a limit in judging whether a deviation is to be considered significant or not. Deviations exceeding twice the standard deviation [under a standard Normal distribution defined on $\mathbb{R}^1$] are thus formally regarded as significant.” He also recast Pearson’s descriptive statistics into a model-based statistical induction, which changed the \textit{ad hoc} approaches before him \citep{spanos2019probability}.

To better understand the \textit{P}-value-based decision-making, it is perhaps helpful to answer the following questions:

\begin{enumerate}
\item What is the fundamental goal of a hypothesis test?
\item What is the statistical model and assumptions used to address the goal?
\item How to formulate the test?
\item What is the difference between the \textit{pre}- and \textit{post-data} views of a hypothesis test? 
\end{enumerate}

\textbf{(1) The fundamental goal.} The fundamental goal of performing hypothesis testing is to derive evidence from the observed data, say $\bx_0 \coloneqq (x_1,x_2, \ldots, x_n)$, to uncover the data generating mechanism M that yields $\bx_0$. More precisely, one wants to learn about the properties of a parameter  $\theta \in \Theta \subset \mathbb{R}^m$ underpinning M. Denoting the true (but unknown) parameter as $\theta^*$, the data generating mechanism can be expressed as \citep{spanos1986statistical}:
\begin{equation*}
\mathcal{M}^* (\bx)={f(\bx; \btheta^*)},   \bx \in \mathbb{R}_X^n
\end{equation*}
where $f(\bx; \btheta^*) $ indicates the joint distribution of $\bx$ with fixed value $\theta^*$ and $\mathbb{R}_{\bX}^n$ denotes the sample space.

Since $\btheta^*$ is unknown (thus $\mathcal{M}^* (\bx)$ unknown), one needs a (realistic) statistical model $\mathcal{M}_{\btheta}^* (\bx)$ (see below) to learn about the true data generating mechanism by estimating the parameter from observed data $\bx_0$. This can be summarized as $\btheta^* 
\stackrel[ \mathcal{M}_{\btheta} (\bx) ]{ \mathcal{M}^* (\bx)  } {\rightleftarrows}
 \bx_0$, where the top arrow represents the data generating mechanism and the bottom arrows indicates parameter evaluation via a statistical model.

\textbf{(2) The statistical model.} A hypothesis test uses a statistical model to learn about the true value of $\btheta$. A general model is:
\begin{equation*}
\mathcal{M}_{\btheta} (\bx) = \left \{ 
f(\bx; \btheta), \btheta \in \Theta \subset \mathbb{R}^m \right \}, \bx \in \mathbb{R}_{\bX}^n
\end{equation*}
where $f(\bx; \btheta)$ denotes the joint distribution of $\bx$ for any given parameter $\btheta$ in parameter space $\Theta$ and $n>m$. 

Thus, the model $\mathcal{M}_{\btheta} (\bx)$ provides a vehicle to perform the hypothesis test (see below) to evaluate the choices of $\btheta$ in $\Theta$ to find an estimate close to $\btheta^*$. Critical to this process are probabilistic modelling assumptions, such as the identically distributed assumption, that underline $\mathcal{M}_{\btheta} (\bx)$. Any violation of the assumptions may yield misleading inference regarding the parameter.

\textbf{(3) The hypothesis test.}  In general, a hypothesis begins supposing that the fixed but unknown parameter locates somewhere in a parameter space. Subsequently, it calculates a (test) statistic from the data ($\Delta(\bx)$) from which tail probabilities (the \textit{P}-value) are calculated to evaluate the strength of support for the hypothesis or lack thereof. In frequentist statistics, no matter how the parameter space is partitioned, $\btheta$ is not a random variable; rather, it is a fixed unknown value (see below). 

\begin{figure*}[h!]
\includegraphics[width=140mm]{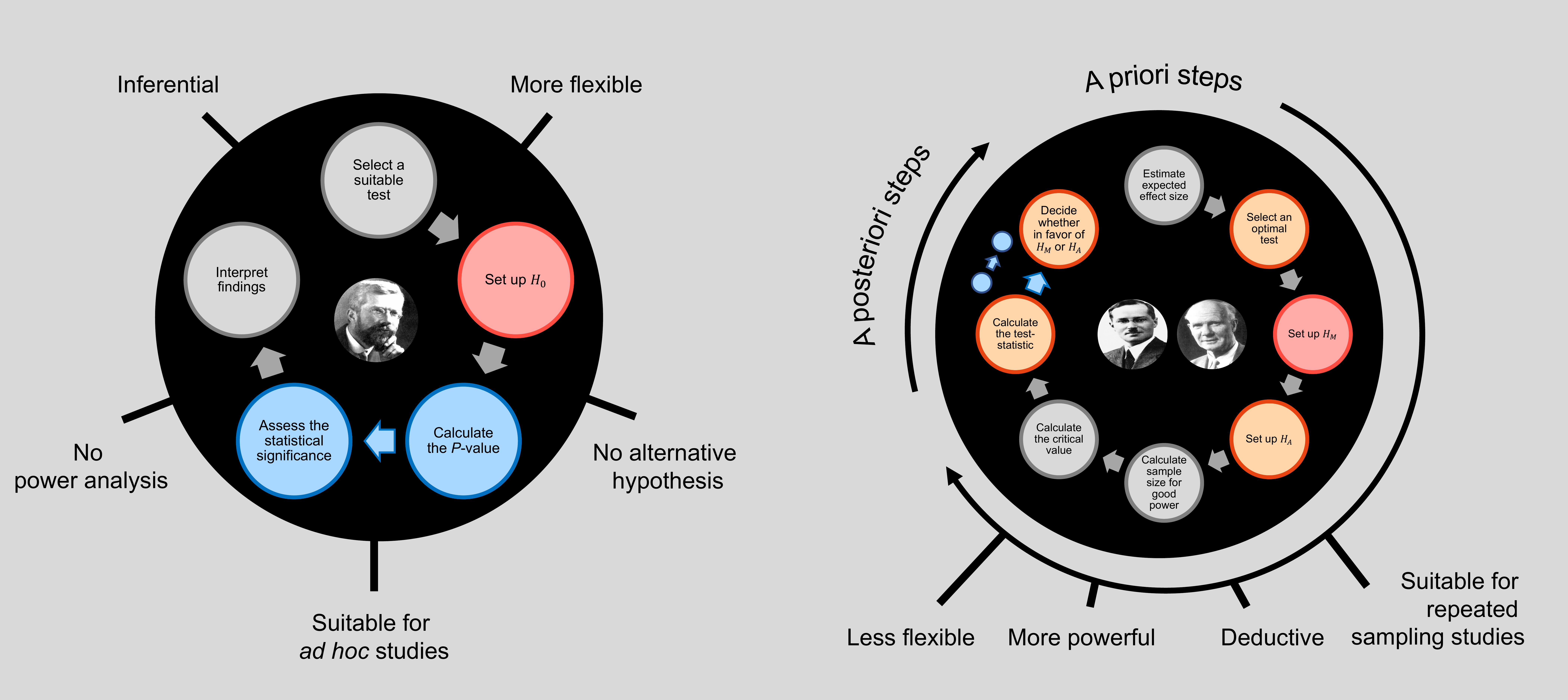}
\caption{A comparison between the Fisher’s hypothesis test and the Newman-Pearson test, and how they constitute the null hypothesis significance test (NHST).}
\smallskip
\parbox[c]{\hsize} {\textbf{Left}: The Fisher’s test by R.A. Fisher. It contains five main steps, following the order clockwise. Compared to the Newman-Pearson test, the Fisher’s test is more flexible, suitable for \textit{ad hoc} studies, inferential, but it does not have an alternative hypothesis nor performs power analysis. \textbf{Right}: The Newman-Person test by Jerzy Neyman and Egon Pearson. It consists of eight main steps, following the order clockwise, where the first six steps are done \textit{a priori}, and the last two steps \textit{a posteriori}. Compared to Fisher’s test, it is more powerful, deductive, suitable for repeated sampling studies, but is less flexible. The null hypothesis significance test is a hybrid of the two; it follows the NP-test procedurally and Fisher philosophically \citep{cortina1997logic, hubbard2004alphabet, johnstone1986tests, perezgonzalez2015fisher, spielman1978statistical}. Specifically, its mandatory steps consist of the steps highlighted in orange in the NP test, with the main hypothesis $H_M$ replaced by $H_0$, and the \textit{P}-value calculation and significance assessment from the Fisher’s test (highlighted in blue) added. 
}
\label{Fisher_vs_NP}
\end{figure*}

Since the NHST\footnote{There are in general three types of hypothesis tests: The Fisher’s test (or the test of significance) \citep{fisher1925statistical, fisher1932inverse, fisher1935design, fisher1955statistical, perezgonzalez2015fisher}, the Newman-Pearson test (or the test of statistical hypotheses or the NP test) \citep{neyman1928use}, and the null hypothesis significance test (or the NHST, a hybrid of the Fisher’s and NP test) \citep{spielman1978statistical, johnstone1986tests, cortina1997logic, hubbard2004alphabet, perezgonzalez2015fisher} (see \textbf{Fig}. \ref{Fisher_vs_NP}). The key difference between the Fisher’s and the NP tests lies in the way the tests treat the parameter space. The former does not partition the parameter space and the latter partitions the space into a null space and an alternative space.}  has been widely used in scientific studies today (\textit{e.g.}, biological studies \citep{lovell2013biological}, education \citep{carver1993case}, psychology \citep{gigerenzer2004mindless, nickerson2000null}, social sciences \citep{Frick1996appropriate}), and has been adopted by textbook writers, journal editors, and publishers \citep{gigerenzer2004mindless, hubbard2004alphabet}, we use the NHST to develop our discussion.

\subsection{The definition of the \textit{P}-value}

Put simply, the \textit{P}-value is the tail probability calculated using the test statistic. To define it formally, let us use an example. 
A psychologist was interested in estimating the average human fluid intelligence (Gf) in a specific age group. Suppose Gf follows a Normal distribution, and we denote $X_i$ as the Gf score for any individual $i \in \mathbb{N}^+$. Then:
\begin{equation*}
X_i \stackrel{iid} \sim N (\mu, \sigma^2) 
\end{equation*}
where $iid$ means independent and identically distributed, with $\mathbb{E}(X_i )=\mu$ and finite $\text{Var}(X_i )=\sigma^2>0$.

Having no prior information about the disease, the psychologist hypothesized that the average intelligence was less than or equal to 100 in that age group. That is, the psychologist hypothesized that the unobserved (true population) mean $\mu$ was less than or equal to $\mu_0$, where $\mu_0$ is set at 100; note that $\mu$ is a fixed value, not a random variable. This forms the null hypothesis $H_0: \mu \leq \mu_0$. In other words, the null hypothesis is true so long as the true parameter falls in the parameter space $M \coloneqq [ 0, \mu_0 ]$. The alternative hypothesis is that $\mu$ was greater than $\mu_0$, namely $H_a: \mu = \mu_1$, for any $\mu_1 > \mu_0$.

Under the \textit{pre-data} view, two useful test-statistics are:
\begin{eqnarray}
\Delta_1(\bX) = \sqrt{n}(\bar{X}_n - \mu_0) / \sigma  
\hspace{2mm}
\stackrel{\mu = \mu_0} \sim 
\hspace{2mm}
N(0,1). \nonumber 
\\
\smallskip
\Delta_2(\bX) = \sqrt{n}(\bar{X}_n - \mu_0) / \sigma 
\hspace{2mm}
\stackrel{\mu = \mu_1} \sim 
\hspace{2mm}
N(\eta,1), \nonumber
\nonumber
\end{eqnarray}

\noindent where $\eta = \sqrt{n}(\mu_1 - \mu_0) / \sigma$, for any $\mu_1 > \mu_0$. 
\smallskip

Let $c_1 (\alpha)= \{ \bx:( \Delta( \bX)>c_\alpha) \}$ be the rejection region. One can define the \textit{P}-value, type I error (or $\alpha$), type II error (or $\beta$), and power (or $1-\beta$) as $p(\bx_0) \coloneqq \mathbb{P}(\Delta_1 (\bX) > \Delta (\bx_0); \mu = \mu_0) $, $\alpha \coloneqq \mathbb{P}(\Delta_1(\bX) > c_\alpha; \mu = \mu_0)$, $\beta \coloneqq 1 - \mathbb{P} (\Delta_2(\bX) > c_\alpha; \mu = \mu_1 )$, and $\mathbb{P} (\Delta_2(\bX) > c_\alpha; \mu = \mu_1) = 1 -\beta$, respectively.

Notice that, the probability of a type II error cannot generally be computed because it depends on the population mean which is unknown. But it can be computed, however, for given values of mean, standard deviation, and sample size. 

\textbf{Remark 1. The parameter is an unknown constant not a random variable in frequentist statistics}. In the frequentist view of the hypothesis testing, the parameter is considered as an unknown constant; not a random variable\footnote{To see this, suppose we can write the conditional density of data $\bx$ given $\theta$,
\begin{equation*}
f( \bx | \theta = \nu) 
= \dfrac{f(\bx | \theta = \nu)}{\int f(\bx | \theta =\nu) d \bx}
\end{equation*}
where $f( \bx | \theta = \nu) $ is a joint density, $\bx \in \mathbb{R}_{\bX}^n$, and $\theta$ is a parameter underlying the statistical model $M_\theta (\bx)$. In frequentist statistic, $\theta$ is some constant that either lays in the null parameter space $\Theta_0 \subset \mathbb{R}^m$ or the alternative parameter space $\Theta_1 \subset \mathbb{R}^m 
\backslash \Theta_0$ (see \textbf{Fig}. \ref{Description_of_P} c). Thus, the joint density $f( \bx | \theta = \nu) $ makes no probabilistic sense (\textit{i.e.}, is not variable with $\theta$). 
} (see \citep{spanos2010frequentist}). Similarly, the \textit{P}-value, the significance level ($\alpha$), and power ($1-\beta$) also do not involve conditioning in frequentist statistics.  

\textbf{Remark 2. The parameter is a random variable in Bayesian statistics}. What if we have some prior information about the parameter $\theta$? For example, we have some (say weak) information about $\theta$, and would like to perform a test to examine whether the data support the null hypothesis, say $H_0: \theta \leq \theta_0$? Said differently, we have some (distributional) information about $\theta$, with which one could already (without seeing the data) form a degree of brief about the hypothesis by evaluating $\mathbb{P}(\theta \leq \theta_0 )$. Then, after seeing data $\bx$, would our brief about $\theta$ be changed?

Formally, suppose we have some prior knowledge that $\theta \sim N (\theta_\pi,\sigma_\pi^2)$. The likelihood of drawing data $\bx = (x_1, x_2, \ldots, x_n)$ is
$\mathbb{P} ( \bx \vert \theta_x )= \prod_{i=1} ^n \mathbb{P} ( x_i \vert \theta_x ) 
= (2 \pi \sigma_x^2 )^{ -n/2} $ 
$\exp \left\{ - \dfrac{1}{2 \sigma_x^2} \sum_ {i=1} ^n  (x_i- \theta_x )^2 \right\}$.
It follows that after seeing data $\bx$, the posterior distribution of  $\theta$ becomes $\theta \vert \bx \sim N (\theta_n, \sigma_n^2)$, where $\theta_n = \left( \dfrac{n}{\sigma_x^2} + \dfrac{1}{\sigma_\pi^2} \right )^{-1} \bigg [
\dfrac{n}{\sigma_x^2} \left (  \dfrac{\sum_{i=1}^n x_i }{n} \right ) 
+ \dfrac{1}{\sigma_\pi^2} \theta_\pi
\bigg ] $ 
and
$\sigma_n^2 = \left (
\dfrac{n}{\sigma_x^2} + \dfrac{1}{\sigma_\pi ^2}
\right ) ^{-1}$. Thus, the evidence for $H_0$ after seeing data is $ \mathbb{P}( \theta  \leq \theta_0  | \bx ) = \mathbb{P} \bigg ( \dfrac{\theta - \theta_n}{\sigma_n} \leq \dfrac{\theta_0 - \theta_n}{\sigma_n}  \bigg \vert \bx  \bigg ) = \Phi \left (\dfrac{\theta_0 - \theta_n}{\sigma_n} \right) $, where $\Phi$ is the $CDF$ for $N(0,1)$.

\begin{figure*}[h!]
\includegraphics[width=143mm]{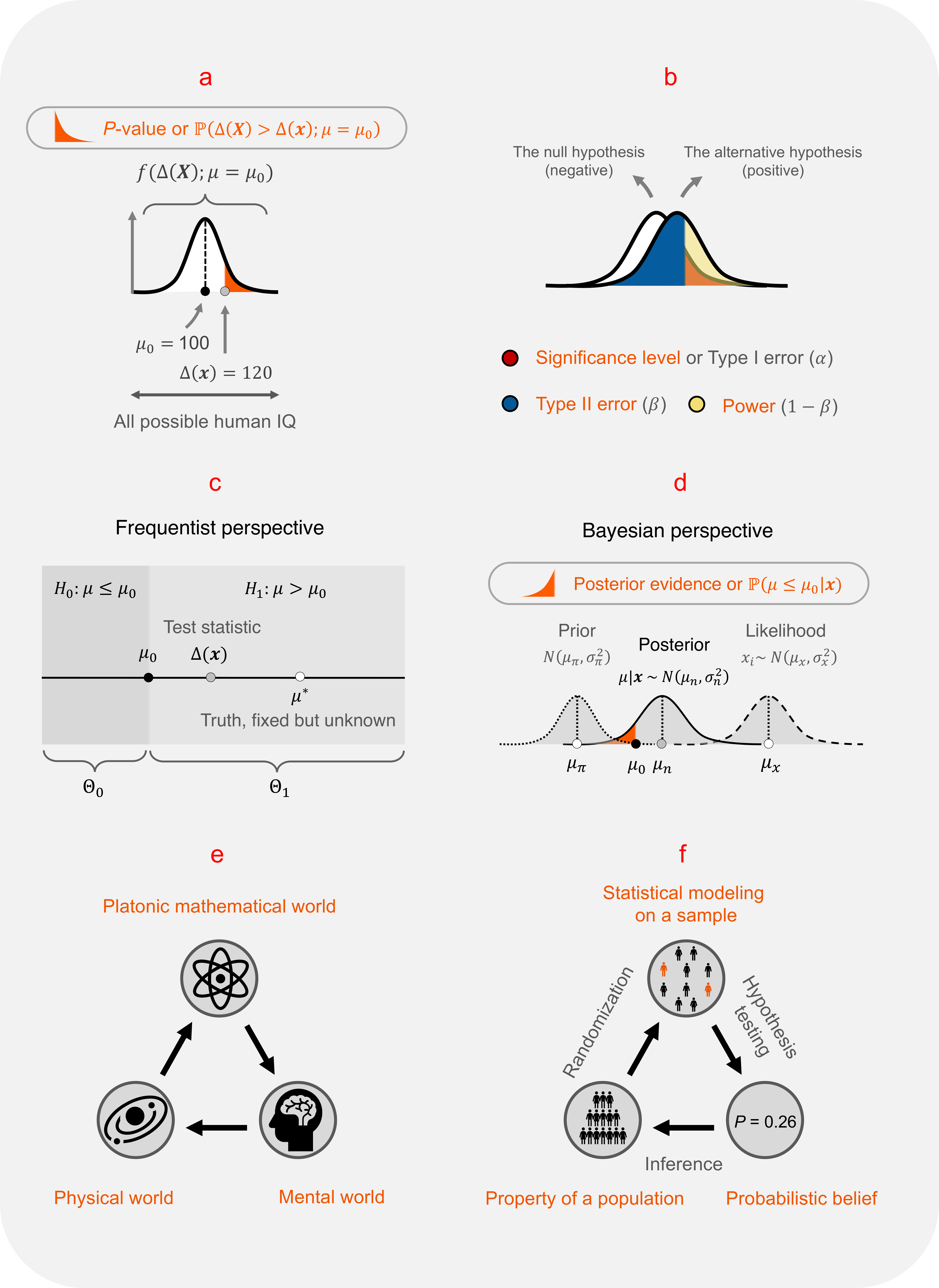}
\caption{The \textit{P}-value and related concepts.}
\smallskip
Caption on the next page.
\label{Description_of_P}
\end{figure*}

\begin{figure*}[h!]
  \contcaption{
  Continued.
   \textbf{(a) Calculating the \textit{P}-value.} (b) \textbf{Significance level (type I error), type II error, and power.} The significance level (type I error or $\alpha$) is a pre-determined value (say $0.05$), which quantifies the probability of observing extreme values given that the null hypothesis were true (red shades). The type II error (or $\beta$) quantifies the probability of failing to reject the null hypothesis given that the alternative hypothesis were true (blue shades). The power (or $1-\beta$) quantifies the probability of rejecting the null hypothesis given that the alternative hypothesis were true (yellow shades). The value of $1-\alpha$ quantifies the probability of failing to reject the null hypothesis when it were true (represented by the white area – not completely shown - under the null hypothesis curve). \textbf{(c) The frequentist perspective of the \textit{P}-value.} In the frequentist view of the hypothesis testing, the parameter is considered as an unknown constant, rather than a random variable. \textbf{(d) The Bayesian perspective of evidence seeking.} Suppose the prior knowledge weakly support for the null hypothesis $H_0: \theta \leq \theta_0$ (with a mean $\theta_\pi$ sits slightly left of $\theta_0$), and the likelihood function has a centre $\theta_x$ that is far right from $\theta_0$. Then, the posterior mean $\theta_n$ is pulled, after seeing the data, in a direction rightwards away from $\theta_\pi$  and towards $\theta_0$ and beyond; the further the centre of the likelihood function is from $\theta_0$ (namely the more evidence the data provide against the null), the further the posterior mean $\theta_n$ is pulled rightward away from $\theta_0$, and there is, therefore, stronger \textit{a posteriori} evidence supporting the alternative hypothesis. \textbf{(e) The three-world system - the physical world, the Platonic mathematical world, and the mental world - and our modification of it.} The physical world represents the entire universe (from every chemical element to every individual) and contains properties that are not readily accessible to the observer. These properties are governed by and can be explained using mathematical principles. The mathematical principles translate into (mental) understanding and form one’s perspective about the physical world. (\textbf{f) The role of the \textit{P}-value in making (causal) inference.} Consider an example where a clinician was enquiring into the prevalence of a disease in a specific age group (\textit{i.e.}, a specific population). Suppose the clinician considered a null hypothesis where the prevalence was 10\% (in the population). Since measuring the prevalence of a disease in a population was impractical, the clinician selected a random sample of ten individuals from the population falling in that age group (see left arrow) and found that two had the disease (see top circle). The clinician then conducted a hypothesis test which generated a \textit{P}-value of $0.26$ (see right arrow) and used this to make inference about the population (see bottom arrow). Given the \textit{P}-value, the clinician concluded that there was not enough evidence (at a significance level of $0.05$) from the sample that would reject the null hypothesis (made about the population). 
  }
\end{figure*}

Figuratively (see \textbf{Fig}. \ref{Description_of_P} d), suppose the prior knowledge weakly support for the null hypothesis $H_0: \theta \leq \theta_0$ (with a mean $\theta_\pi$ sits slightly left of $\theta_0$), and the likelihood function has a centre $\theta_x$ that is far right from $\theta_0$. Then, the posterior mean $\theta_n$ is pulled, after seeing the data, in a direction rightwards away from $\theta_\pi$ and towards $\theta_0$ and beyond; the further the centre of the likelihood function is from $\theta_0$ (namely the more evidence the data provides against the null), the further the posterior mean $\theta_n$ is pulled rightward away from $\theta_0$, and there is, therefore, stronger \textit{a posteriori} evidence supporting the alternative hypothesis.  

To avoid confusion, when we speak of\textit{ P}-value below, we refer to it in the frequentist sense. We will discuss Bayesian evidence in \textbf{Section} 5. 

\textbf{Remark 3. The \textit{P}-value represents \textit{post-data} evidence and is arguably inappropriate in a two-sided test}. Type I error, type II error, and power are \textit{pre-data} probabilities (or evidence) \citep{spanos2013should}. In other words, they do not involve observations $\bx=(x_1,x_2, \ldots,$ $x_n)$. Practically, one sets the experimental conditions (for example, by choosing a specific sample size and the experimental mechanism during a clinic trial) such that the experiment yields pre-specified levels of type I error, type II error, and power. As such, these probabilities are embedded in the experiential design before the data have ever been seen. 

The \textit{P}-value, by involving the observation $\bx$ (in the test statistic $\Delta(\bx)$), is a \textit{post-data} error probability (see Ch.13 in \citep{spanos2019probability}). In other words, one needs to see the data $\bx$ first and then calculate a \textit{P}-value; the former affects the latter. This is also true for Bayesian evidence, as the posterior depends on both the prior and the likelihood (the latter of which depends on the data); we will discuss about this further in \textbf{Section} 5.

A two-sided test is arguably not suitable under the \textit{post-data} view. To see this, suppose that one is testing whether the average human Gf is 100, \textit{i.e.}, $H_0: \mu_0=100$ \textit{vs.} $H_1: \mu_1 \neq 100$. Suppose the average Gf score from a randomized, representative sample is $110$, then, \textit{post-data}, one already knows that it lays one the right-hand side of $\mu_0$ and thus one only needs to calculate the tail probability on the right-hand side. 

Naturally, one may ask, what if I draw another sample and found that the average Gf score from the second sample is $90$, which yields the following contradiction where the \textit{P}-values from Samples 1 and 2 may each reject the null but after combining the two samples, the \textit{P}-value may fail to reject the null. It is also possible that the combined sample would still reject the null but the \textit{P}-value changes. Regardless of the scenarios, the key point is that the \textit{P}-value provides \textit{post-data} evidence and may be affected by the nature of the (sample) data.

To summarize: 
\begin{itemize}
\item The \textit{post-data} view of the \textit{P}-value suggests that considering two-sided tail probabilities for each sample may be redundant after seeing the observations.
\item The \textit{P}-value provides data-specific evidence. This does not suggest that the \textit{P}-value is useless or wrong. The heart of a statistical enquiry is to discover knowledge from the data (we have at hand) via a logical argument (\textit{e.g.}, hypothesis testing); the \textit{P}-value offers a simple, convenient, and perhaps universally applicable probabilistic measure to summarize information and helps to draw inference or conclusion from the data. 

\item When the sample is representative or significant results have been reproduced in different samples and studies, the message (\textit{e.g.}, a significant finding) delivered by the \textit{P}-value would be stronger and potentially more reliable (than it from a study with a non-representative sample or one yet to be reproduced). In general, however, when interpreting hypothesis-testing based evidence, one needs to bear in mind that the conclusion drawn may be sample-specific, and the \textit{P}-value may be affected by how the sample is collected, how data are aggregated and processed, what the sample size is, \textit{etc}. This calls for the use of the \textit{P}-value in context (see \textbf{Section} 4).
\end{itemize}

\section{The interpretation of the \textit{P}-value in decision-making}

\subsection{The philosophy of the \textit{P}-value in decision-making}
In our view, a hypothesis testing framework links the population (\textit{e.g.}, a group of individuals), a statistical model, and some (mental) probabilistic belief\footnote{This is inspired by Roger Penrose’ three-world system linking the physical, mathematical, and metal worlds \citep{penrose2005road}.}. The population has interesting properties (\textit{e.g.}, the prevalence of a disease in the population) governed by some data generating principle that is difficult to state explicitly. To gain insights about a particular property, one develops a hypothesis about the data generating principle (for example, that the underlying genetic and environmental factors may yield a prevalence of 20\%). To evaluate this hypothesis, one then draws a sample (in a proper manner) and test whether there is evidence for it (see left arrow in \textbf{Fig}. \ref{Description_of_P} (f)). The test produces a \textit{P}-value (right arrow in \textbf{Fig}. \ref{Description_of_P} (f)), with which one assigns probabilistic belief about the hypothesis (bottom arrow in \textbf{Fig}. \ref{Description_of_P} (f)) and concludes whether to reject it or not. 

Such a system must confront a few flaws. First, it may be possible that the sample property does not well represent the population property. Next, the unknown property of the population may not be well established using a statistical argument (\textit{e.g.}, a test done on a sample whose distribution violates the assumption of the test). Thus, the \textit{P}-value and the belief attached to it (to make any statement about the population property) via a hypothesis test may be inconsistent with the true (but unknown) population property. 

\subsection{The Roles of the \textit{P}-value in Scientific Enquires}

In spite of criticisms\footnote{“… it does not tell us what we want to know, and we so much want to know what we want to know that, out of desperation, we nevertheless believe that it does” \citep{cohen1994earth}.},  the \textit{P}-value has been of great interest to statisticians, biological and medical scientists, clinicians, and philosophers in its three-century long history \citep{skipper1967sacredness, panagiotakos2008value,singh2008interpreting,  richard2017statistical}.

Hypothesis testing and the \textit{P}-value form a knowledge-acquiring system that derives evidence from a sample; they also form an inferential system that throws probabilistic light on the population. There are, in general, four important roles that the \textit{P}-value plays in scientific enquires. First, it allows for comparing and bridging decision-making outcomes regarding the same testing problem done on different studies and datasets\footnote{“Different individuals faced with the same testing problem may have different criteria of size (see effect size in the enclosed glossary and abbreviations, our insertion). Experimenter I may be satisfied to reject the hypothesis $H$ using a test with size $0.05$, whereas experimenter II insists on using $0.01$. It is then possible that experimenter I rejects $H$, whereas experimenter II accepts $H$ on the basis of the same outcome of an experiment. If the two experimenters can agree on a common test statistic, this difficulty may be overcome by reporting the outcome of the experiment in terms of the \textit{P}-value.” (see \citep{bickel2015mathematical}, p. 221).}. Second, it supports evidence at a continuous scale (rather than binary conclusions)\footnote{“… the smaller the \textit{P}-value, the stronger the evidence for rejecting the null hypothesis. Hence, a \textit{P}-value reports the results of a test on a more continuous scale, rather than just the dichotomous decision ‘Accept the null hypothesis’ or ‘Reject the null hypothesis’” (See \citep{casella2021statistical}, p 397).}; but see\footnote{The \textit{P}-value is not just data dependent but also model-based; it cannot be properly interpreted out of its statistical context \citep{spanos2019probability}.}. Third, it enables integrating results from multiple studies and datasets\footnote{When different experiments produce various types of data, the \textit{P}-value can combine the evidence relating to a given hypothesis \citep{van1967combination}. This is the basis for ‘data fusion’ and meta-analysis \citep{hedges1985statistical}.} (see more in \textbf{Section} 5). Fourth, it facilitates causal inference. 

\subsection{The \textit{P}-value in Hypothesis Testing and Causal Inference}

Aristotle said, “We do not have knowledge of a thing until we have grasped its why; that is to say, its cause.” The contributions of hypothesis testing and the \textit{P}-value to causal studies, in general, lie in five areas: the estimation of a causal effect, cross-validation (including out-of-sample testing), graphical causal reasoning, cause alteration, and the method of \textit{instrumental variables} (IV). 

We discuss them in order below using examples in brain studies about which we know slightly more. Although the applications of causal inference may differ from one subject to another, the general roles of the \textit{P}-value suggest that there is a common theme which one can glimpse into by focusing on one subject. Certainly, we do not imply that there are no other contributions they make to causal inference. But if there are, it would be difficult to list every departure or derivative of it. We hope that our presentation may stir further discussion and that our ever-expanding statistical and scientific knowledge will one day allow us to formulate more universal statements about hypothesis testing and the \textit{P}-value.

\begin{figure*}[h!]
\includegraphics[width=140mm]{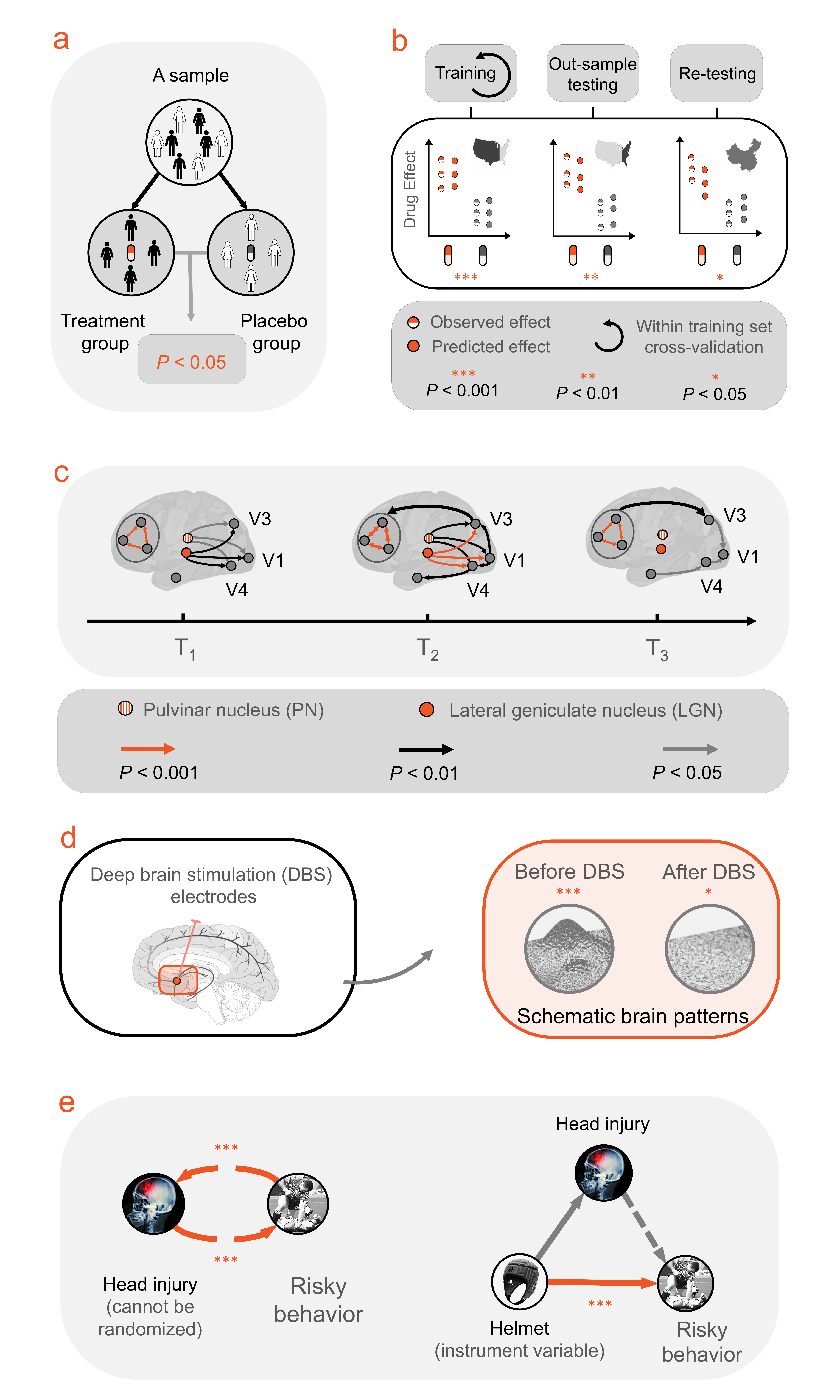}
\caption{The roles of the hypothesis testing and the \textit{P}-value in making causal enquiries.}
\smallskip
Caption on the next page.
\label{Causal_inference}
\end{figure*}

\begin{figure*}[h!]
  \contcaption{
  Continued.
 \textbf{(a) The estimation of a causal effect.} The average causal effect in a randomized study can be identified and quantified using the difference between the expected outcome of the treatment group and the control group, and subsequently examined via a \textit{P}-value. \textbf{(b) Out-of-sample test.} The model performance or the causal effect estimated from one dataset, if not validated, may be exaggerated or overfit the dataset. Out-of-sample testing can, to a certain degree, alleviate over-fitting by training the model using a subset of the data (left) and testing it in the remaining, previously unseen, data (middle). Additional testing using data from another study or demographically distinctive sample may further support the generalization of the trained model and its suggested causal claims (right). \textit{P}-value is critical to evaluate whether the tests are successful, thereby guarding their validity and efficacy. \textbf{(c) Graphical causal reasoning.} The directed arrows (or edges) indicate potential causation. The figure gives an example of directed causal flows in the brain when performing object recognition. When one views an object, areas in the visual cortex including V1, V3, and V4 first receive inputs from pulvinar nucleus (PN) and lateral geniculate nucleus (LGN) (left figure). Subsequently, V1 sends signals to V3 (which processes object recognition) and V4 (which processes colour recognition), and through V3, sends information to the prefrontal cortex (middle figure). Finally, there are reverse feedbacks from V3 and V4 to V1 (right figure). \textbf{(d) Causal alternation.} If altering the stimuli (while controlling for covariates) results in a change of the outcome, then it suggests that the former cause the latter. The figure gives an example of  the deep brain stimulation (DBS), where applying DBS to a target brain region, the brain patterns of the area change accordingly, which then modifies (behavioural) symptoms. DBS is used in treating severe Parkinson’s disease.  \textbf{(e) The method of \textit{instrumental variables} (IV).} When directly altering causes or randomization is unavailable, one can consider the method of IV. In the figure, one is interested in studying whether a head injury causes risky behaviour. On the one hand, randomization or assigning a head injury is impossible; on the other hand, it could be argued that reverse causation, where risky behaviour causes a head injury, is also possible. By using an IV, wearing a helmet, one can then study whether a head injury causes risky behaviour. Suppose one assumes that wearing a helmet is unlikely to cause risky behaviour (in the long term), and it is likely to reduce a head injury. If introducing wearing helmets reduces risky behaviour (while controlling for all other variables such as age and gender), it suggests that wearing helmets reduces a head injury, which reduces risky behaviour. 
  }
\end{figure*}

First, the \textit{P}-value helps to \textbf{estimate a causal effect} (see \textbf{Fig}. \ref{Causal_inference} (a)). Suppose a researcher is interested in studying whether a Levodopa-based drug is effective in treating Parkinson’s disease (PD). They need to compare the symptoms of a PD patient after taking the drug to that of the \textit{same} (our emphasis) patient not taking the drug. Only one of the two is observable and within-subject designs are not suitable due to carry-over effects. Using randomization
\footnote{There are times when randomization becomes impossible. For example, it is unethical to assign a group of 45-year-old healthy subjects to take a new Levodopa-based drug to investigate whether the drug reduces one’s PD symptoms at 50. Additionally, there is likely another source, say, the socioeconomic status (which may be related to the affordability of new drugs) or genetics (if there is a family history of PD, one may be more willing to take the drug), that is both associated with taking the drug and developing PD at 50. Similarly, it would be difficult to estimate the effect of taking the drug on reducing PD symptoms by comparing the PD symptoms of an individual at 50 who had taken the drug with his or her PD symptoms at 50 had he or she not taken the drug. To solve these issues, Propensity Score Matching (PSM) estimates the treatment effect by comparing the outcomes of the subjects under treatment (\textit{e.g.}, taking the drug) with a set of “matched” subjects without treatment (\textit{e.g.}, having not taken the drug) \citep{rosenbaum1983central, caliendo2008some, dehejia1999causal, dehejia2002propensity}. More concretely, one could first compute the propensity score of A’s taking the drug based on his or her gender, economic, social, genetic, and demographic backgrounds, and choose an individual from a group of 50-year-old who had not taken the drug but has a propensity score (of taking the drug during his or her younger years) closest to A’s. Then we can compare the PD symptoms between these two individuals and estimate the effect of taking the drug on reducing PD symptoms at age 50.}
, the Neyman–Rubin causal model (or the potential outcomes framework) shows that the average causal effect can be identified and estimated using the difference between the expected outcome of the treatment group and the expected outcome of the control group (without randomization, one cannot derive causal properties from two groups consisting of different individuals) \citep{Neyman1923application, neyman1935statistical, rubin1978bayesian}. By evaluating the \textit{P}-value, a hypothesis test can then examine whether, and, if so, to what extent, the drug effect from the treatment group is more significant than that of the control group.

Second, the \textit{P}-value facilitates \textbf{out-of-sample testing}. The \textit{P}-value is useful to verify whether evidence (\textit{e.g.}, hypothesis testing conclusions and model performance) discovered in a sample can be extrapolated to another independent sample (see \textbf{Fig}. \ref{Causal_inference} (b)). For example, if one is interested in developing a model to select neural markers that can predict the severity of Parkinson’s disease (say the MDS-UPDRS score), one can first fit the model on brain data obtained from a training sample of 70 people during model development. Subsequently, one can test the parameters of the trained model on 30 new subjects to check whether the markers can predict the MDS-UPDRS scores in new subjects, without further modelling. The efficacy of the selected neural markers can be evaluated by comparing how well the predictions are made using a distance measure (\textit{e.g.}, Pearson correlation) and its \textit{P}-value. If significant, one can say that the model fitted on the training set is reproducible (with regards to the test set). Additionally, the \textit{P}-value can be used to test
\footnote{Neither type of out-of-sample tests, strictly speaking, examines causation; an out-of-sample study endorsed by \textit{P}-value, however, reduces the likelihood of model overfitting. Although an overfit model suggests nothing about causation (but about association), a reproducible model does offer stronger evidence of association and suggests that the association relationship may be more likely to be causal. In short, out-of-sample testing potentially yields more rigorous statistical claims about model performance, and potential causal relationships between variables under investigation. Overall, when significant results are discovered from an experiment, it is useful to repeat the experiment to verify if the result can be replicated or reproduced \citep{vaux2012replicates}.} 
whether model trained (and results obtained) from one study (including within-study training and testing) can be extrapolated to or reproduced in another dataset or study \citep{cao2018cerebello}. 

Third, the \textit{P}-value is useful in graphical \textbf{causal reasoning}
\footnote{Its modern development is based on Reichenbach’s macro statistical theory \citep{reichenbach1956direction} and Suppes’ probabilistic theory \citep{suppes1970probabilistic} (Interested readers could refer to the books edited by Sosa and Tooley for a thorough discussion \citep{sosa1975causation, sosa1993causation}). }. 
Suppose one uses a graphical model to study how activities from brain region $A$ may be causing those from region $B$ (see \textbf{Fig}. \ref{Causal_inference} (c)). One can perform a hypothesis test and use the \textit{P}-value to evaluate whether a significant \textit{directed} edge exists from $A$ to $B$ (or from $B$ to $A$) \citep{hinton2005kind, pearl1993comment, greenland1999confounding}.

Fourth, the \textit{P}-value is useful to study \textbf{causal alteration}. It examines if modification of a hypothesised cause (while fixing other potential causes unaltered) results in a change of the hypothesised effect (see \textbf{Fig}. \ref{Causal_inference} (d)). For example, via transcranial magnetic stimulation (TMS), one can use a magnetic field coil to generate electric current, which modifies the magnetic field of a specific group of neurons in a small surface region of the brain \citep{lipton2010transcranial, romei2012causal}. After controlling for confounds, one can perform a hypothesis test to examine whether there is a significant difference between the outcomes (\textit{e.g.}, human behaviour) when these neurons are “on” with the outcomes when they are “off” and conclude, based on the \textit{P}-value, whether these neurons are responsible for the outcome change.

When a direct manipulation of the cause is impractical, the \textit{P}-value is useful when employing \textbf{the method of \textit{instrumental variable}} (IV)
\footnote{A suitable \textit{instrumental variable} (IV) is one that is correlated with an endogenous explanatory variable, such as the severity of a head injury, but is not correlated with the error term (for example, in a regression). An endogenous explanatory variable is a covariate that is correlated with the error term.}
(see \textbf{Fig}. \ref{Causal_inference} (e)) \citep{angrist1996identification}. For example, head injury for rugby players may cause behaviour, emotional, and sensory changes (such as developing risky behaviour, becoming irritable and angry, and having trouble with balance). A significant correlation between the severity of head injuries and changes in behaviour, emotional, and sensation, however, does not conclude the former causes the latter. On the contrary, having risky behaviour and being irritable and angry may result in fights between players whereas having a poor sense of balance may cause falling, both of which may result in head injuries. Furthermore, a head injury may first affect another variable, such as developing depression, which then affects the behavioural, emotional, and sensory changes. One cannot randomize individuals to receive a head injury, but could relatively easily introduce an additional variable, or \textit{instrumental variable} (or IV), which affects the chance of having a head injury, but has no independent effect on the outcome (\textit{i.e.}, the behavioural, emotional, and sensory changes). A possible IV here is wearing helmets (in Rugby Union, players usually do not wear helmets), which may reduce the chance of having a head injury but does not directly affect the outcomes. If, after introducing the helmet, the behavioural, emotional, and sensory changes become insignificant, one can conclude with more confidence that head injuries are the cause for changes. The \textit{P}-value helps to evaluate the effect size, strength, and direction of the causal effect of the IV.  

\section{Paradoxes and Potential Misuses of the \textit{P}-value}

In this section, we discuss the paradoxes and misuses of the \textit{P}-value. \textbf{Section} 4.1 enquiries into the relationships between the \textit{P}-value, sample size, and significance level, in hypothesis testing and decision-making. \textbf{Section} 4.2 compares statistical significance and clinical relevance. \textbf{Section} 4.3 presents common misuses and misinterpretations of \textit{P}-value in scientific studies, accompanied by our modest recommendations. We hope that our discussions and suggestions, by no means exhaustive, may improve the use of the \textit{P}-value to deliver more consistent and reproducible scientific discoveries.

\subsection{The Paradoxes of the \textit{P}-value}

Suppose a clinician wanted to test whether the prevalence of a disease was 10\%. To do so, the clinician selected a sample of 10 individuals and found that two out of the 10 had the disease. She then used evidence from the sample (20\% sample incident rate) to make inference about the population prevalence. With \textit{P} $= 0.26$, the hypothesis was not rejected. 

The first paradox of the \textit{P}-value is that decisions made on the same effect size from data of different sizes may be inconsistent. For example, suppose we increased the sample size from 10 to 50, of which 10 had the disorder (the sample incident rate remained at 20\%). This yielded a \textit{P}-value of 0.02. Although the new sample had the same (20\%) incident rate, the null hypothesis was rejected under a significance level of 0.05. This test, however, would still survive under a significance level of 0.005. Now consider an even larger sample of 100, of which 20 had the disease (the sample incident rate remained 20\%), but the \textit{P}-value was 0.002. The hypothesis was rejected under 0.005. 

\begin{figure*}[h!]
\includegraphics[width=140mm]{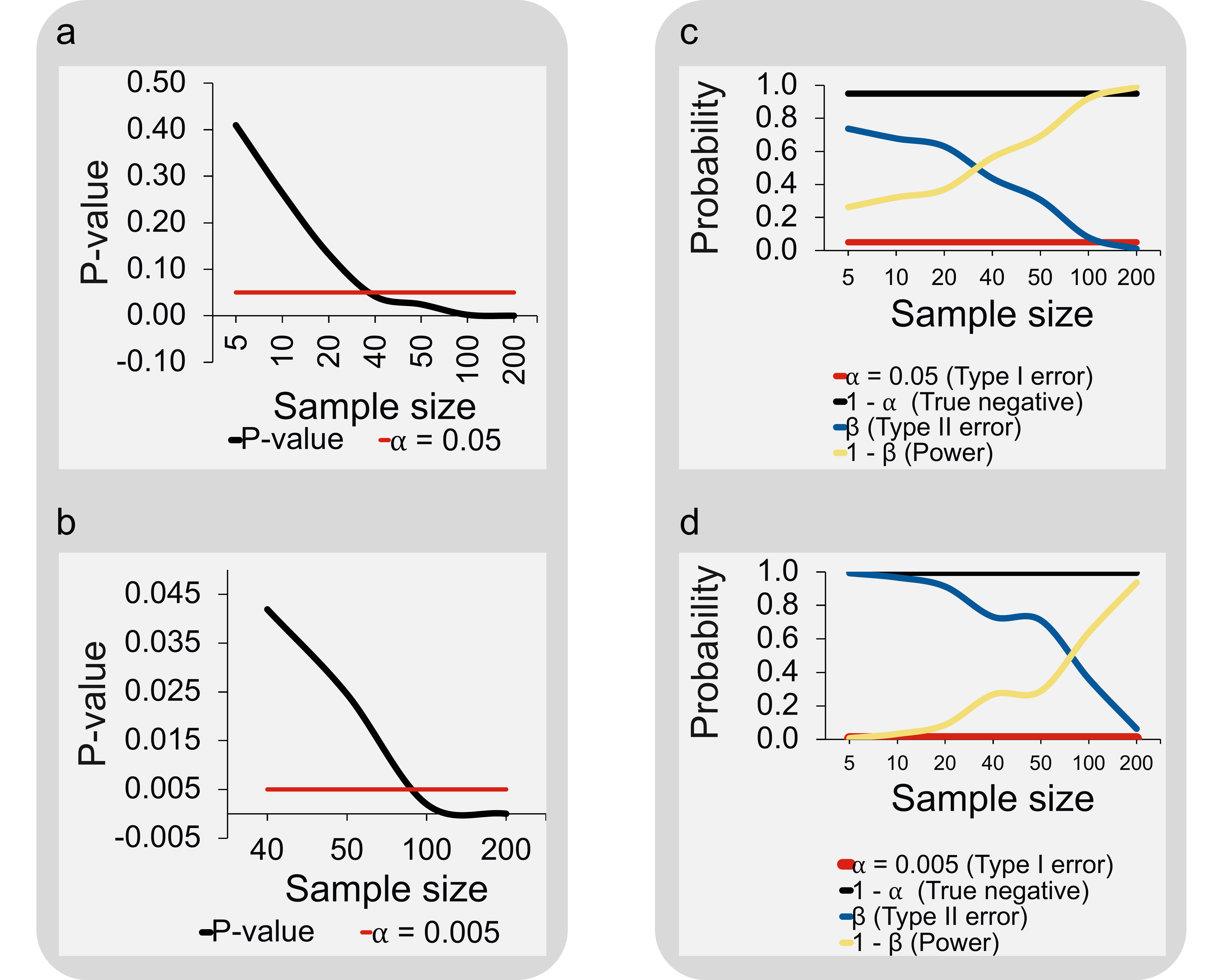}
\caption{The paradoxes of the \textit{P}-value.}
\medskip
\parbox[c]{\hsize} {\textbf{(a)} The associations between the \textit{P}-value, the sample size, and the significance level. The figure shows that the \textit{P}-value goes down as the sample sizes increases. The paradox lies in that, given a particular significance level (say $0.05$), one can increase the size of the sample to obtain a \textit{P}-value that is significant. \textbf{(b)} Even if the significance level is lowered (to, say, $0.005$), one could keep increasing the sample size to obtain a significant \textit{P}-value. On the other hand, with a fixed sample size, one may adjust the significance level to “control” whether the result is significant or not. \textbf{(c)} The paradox between the \textit{P}-value, the sample size, and statistical power. A larger sample size may yield a more significant \textit{P}-value with a small effect size, but it also increases power. \textbf{(d)} Meanwhile, reducing the significance level (say from $0.05$ to $0.005$) may produce more conservative testing results, but it reduces power. \textbf{Figs}. (a)-(d) give demonstrations, from different perspectives, why the interpretation of the \textit{P}-value needs to be contextual.
}
\label{Lady_tasting_tea}
\end{figure*}

\begin{figure*}[h!]
\begin{subfigure}{.5\textwidth}
  \centering
  \includegraphics[width=.95\linewidth]{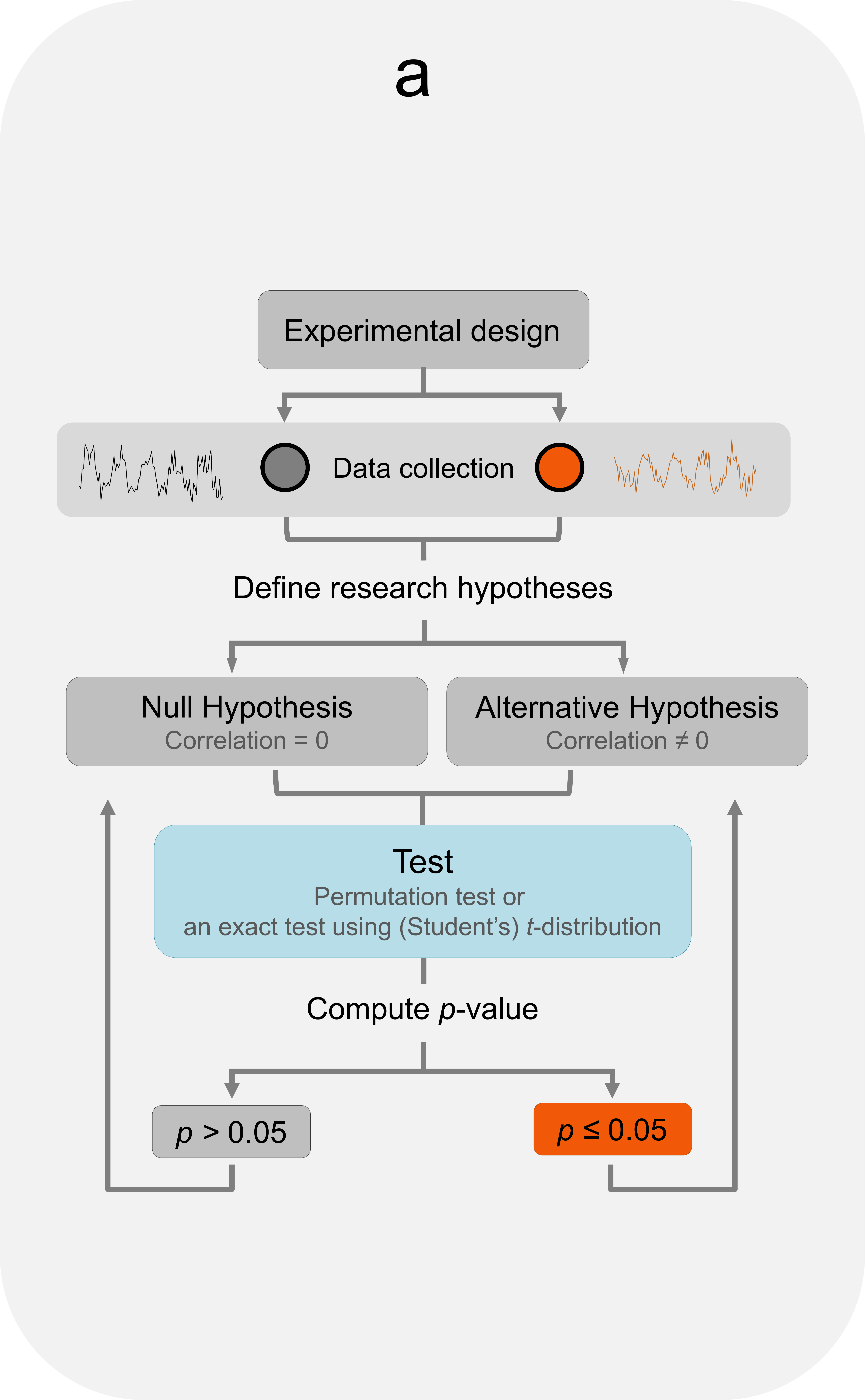}
\end{subfigure}%
\begin{subfigure}{.5\textwidth}
  \centering
  \includegraphics[width=.95\linewidth]{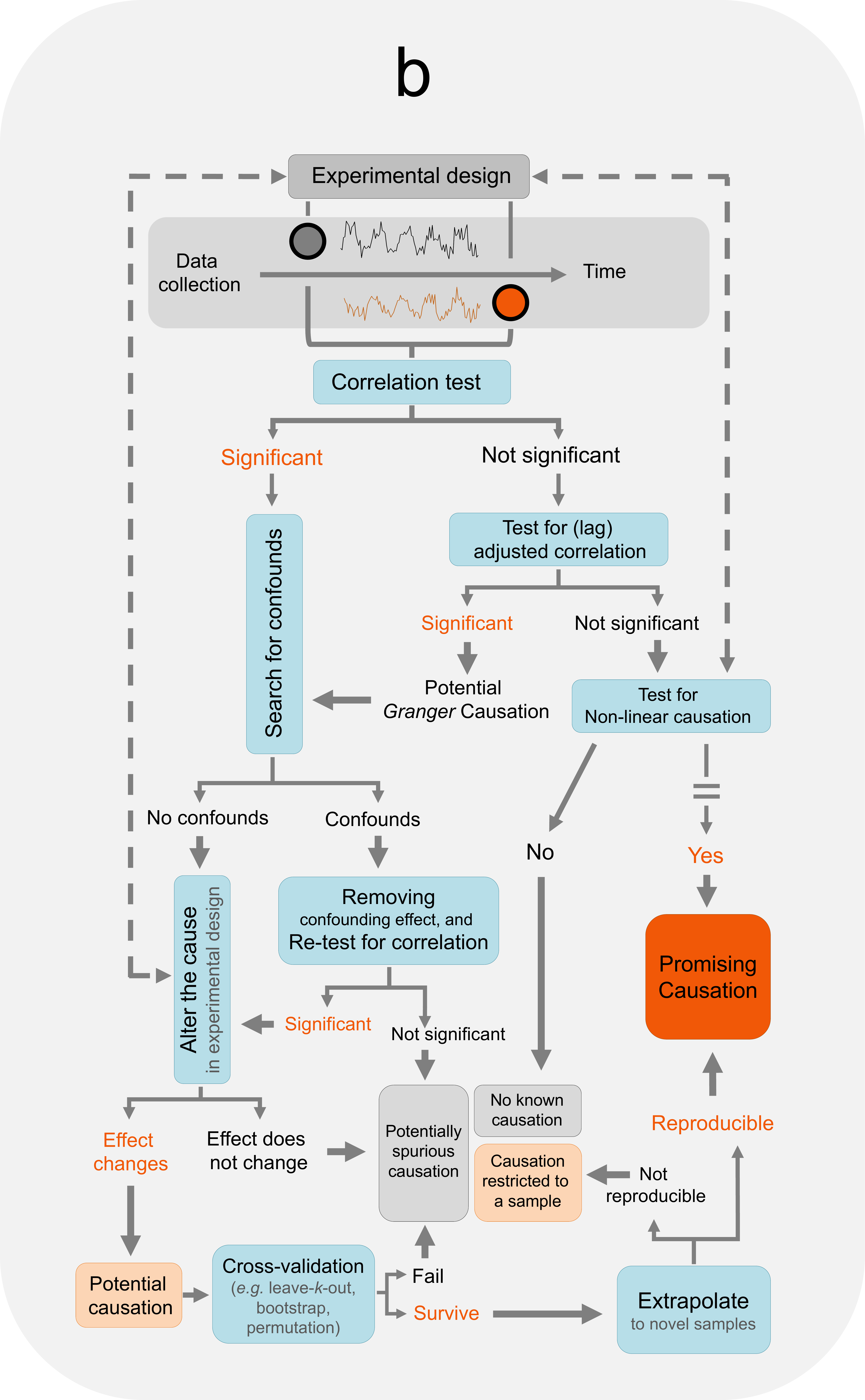}
\end{subfigure}
\caption{Making better use of the \textit{P}-value.}
\medskip
\parbox[c]{\hsize}{\textbf{(a) A typical flowchart for conducting hypothesis-led testing} on, for example, whether the correlation between two random variables is significantly different from zero. A significant correlation, however, does not equate to causation. Note that this framework forms the first part of the flowchart in figure (b). \textbf{(b) A more rigorous flowchart.} We use the correlation test as an example; it could be replaced with other models or tests. It can also extend to cases involving more than two variables. For demonstration, we focus on testing linear causation, and abbreviate the procedure for testing non-linear causation (which is marked with two parallel bars in the figure) - interested readers can refer to \citep{bai2010multivariate, hiemstra1994testing}. We do not claim nor advocate that this is the only procedure to perform hypothesis testing; rather, we show that it is important to remove confounding effects, avoid over-fitting, and conduct reproducible research. A careful experimental design, appropriate data processing, and contextual scientific interpretation are equally important, but are not shown in the figure. The illustration demonstrates that even simple analysis needs additional caution when causal inference and reproducibility are concerned. 
}
\label{Pipeline}
\end{figure*}

Generally, as we see from \textbf{Fig}. \ref{Lady_tasting_tea}, the \textit{P}-value decreases monotonically as the sample size increases
\footnote{This was first observed by \citep{berkson1938some}.}. 
Thus, a hypothetically aggressive scientist may attempt to “hack” the \textit{P}-value by adding more subjects to the study or by repeating significance tests. To avoid this, one may consider both sample size and effect size during experimental plans. For example, in clinical trials, a Phase II study is first done to determine effect size and population variation, and this information is then used to power a Phase III study to ensure collecting enough sample to detect the difference. Indeed, given unlimited resources, most people may prefer studies with very large sample sizes, as they feel larger sample studies are more reliable than smaller trials. Here, we do not advocate against large-sample studies (which have many advantages as we see below); rather, we argue that one should treat the \textit{P}-value contextually and avoid being that aggressive scientist (see suggested guidelines in \textbf{Table} 1 and \textbf{Fig}. \ref{Pipeline}).

The second paradox arises because of the relationship between the sample-size, the \textit{P}-value, and power. To see this, let’s return to the example above and consider a null hypothesis where the prevalence of the disease is 10\% against an alternative hypothesis where the prevalence is 20\%. On the one hand, adding more data may appear \textit{P}-hacking but it improves power: under the same significance level (say $0.05$), the type II error decreases as the sample size goes up; as a result, the power increases. On the other hand, a stringent significance level is not always beneficial: comparing \textbf{Fig}. \ref{Lady_tasting_tea} (c) with \textbf{Fig}. \ref{Lady_tasting_tea} (d), a test with a more stringent significance level yields less power, and this is true for every sample size. 

Taken together, although the incidences in the three samples ($n=10$, $50$, and $100$) were the same, the hypothesis testing results were different. In other words, for each (lower) significance level, when the sample incidence rate was relatively stable, it was possible to obtain a significant \textit{P}-value by increasing the sample size, thereby “hacking” the test. 

These highlight that the interpretation of a \textit{P}-value needs to be contextual. Moreover, when designing experiments and conducting hypothesis testing, there is a compromise to make, one that considers balancing the sample size, significance level, and power.

To summarize:
\begin{enumerate} [(i)]
\item The \textit{P}-value-based hypothesis testing is sample-size dependent.
\item Lowering the threshold alone may make rejecting a null hypothesis more difficult, but one may increase the sample size to “hack” the \textit{P}-value.
\item Increasing the sample size may yield a more significant, but not necessarily meaningful, \textit{P}-value (see \textbf{Section} 3.6), but it increases power. Reducing the significance level (say from $0.05$ to $0.005$) may produce more conservative testing results but reduces power. 
\item The interpretation of the \textit{P}-value needs to be contextual, accounting for the experimental design, model specification, sample size, significance level, the desired power, and the scientific question. 
\end{enumerate}

\subsection{Statistical Significance (\textit{P} $< 0.0X$) \textit{vs}.  Clinical Relevance}

The paradoxes of the \textit{P}-value raise the need to distinguish statistical significance and clinical relevance\footnote{J.D. wrote Section 4.2.}. First, a significant \textit{P}-value may not guarantee clinical relevance. Second, ignoring a statistically non-significant clinical finding (due to, say, high variation in the sample or small sample size) may neglect useful information or yield publication biases \citep{easterbrook1991publication, greenfield1998statistics}. When employing hypothesis tests in clinical studies, one cannot stress enough that having a \textit{P}-value $> 0.05$ only indicates a lack of evidence to reject the null hypothesis; it is not equivalent to ``no difference between groups''. In particular, Mayo and Spanos used the concept of \textit{post-data} severity evaluation to explain how the \textit{P}-value based decision making can cause the fallacies of acceptance (when no evidence against $H_0$ is mistreated as evidence for it given low test power and small sample size in detecting reasonable discrepancies) and rejection (when evidence against $H_0$ is interpreted as evidence for a particular $H_1$ given high test power and a large sample for potentially trivial discrepancies) \citep{mayo2006severe}. The fallacy of rejection concerns a dangerous practice; that is, to conflate statistical significance with substantive (or clinical, for most medical research questions) significance, or, to be more specific, to conflate the statistical alternative with a substantive theory \citep{mayo2006severe}.  The null and alternative hypotheses under Neyman-Pearson's framework must exhaust the parameter space of a given statistical model and thus only allow the alternative hypothesis to be deduced upon the null is rejected, but not based on a substantive theory or knowledge. And a statistically significant effect (\textit{i.e.}, $\Delta(X) > c_\alpha$) needs not to be indicative of a large or meaningful effect size. 

In clinical trials and drug development, clinical significance typically refers to the magnitude of the actual treatment effects, which suggests whether the results of, say, a trial, can impact current medical practice. In other words, a `clinical significance' may be a cardinal element in driving treatment decisions \citep{ranganathan2015common}. That is why hypothesis testing-based \textit{P}-value alone may oversimplify a clinical question or provide insufficient information regarding the clinical (trial) results. As such, information on ``minimal clinically important differences (MCID)'' or ``minimal important changes (MIC)'' needs to be discussed beforehand based on prior knowledge or experiments. Model interpretation in clinical studies, therefore, need combined expertise from statisticians, clinicians, and general scientists. In addition to evaluating the \textit{P}-value, it is useful to take the effect size and the direction of the effect into consideration \citep{du2009confidence}. Suggestions provided in \textbf{Table} 1 may be useful in this regard.

Another way to avoid making decisions based only on the \textit{P}-value is to report both the \textit{P}-value and a confidence interval (CI) \citep{rothman1978show}. Although there is a mathematical duality between obtaining a CI and performing hypothesis testing, the CIs are less vulnerable to the large $n$ problem and are more informative than the \textit{P}-values \citep{mayo2018statistical}. Other advantages to include the CI are: CI (1) reports results directly on the scale of the data, (2) provides the direction and strength of the effects, (3) partly implies sample size and variability through its width \citep{gardner1986confidence, du2009confidence, shakespeare2001improving}, and (4) avoids the problem of sharp dichotomy (\textit{e.g.}, rejecting null at $0.05$ but failing to do so at $0.0499$) \citep{spanos2019probability}. By presenting the CI along with the \textit{P}-value, one may steer away from purely seeking statistical significance, and into considering statistical significance in light of clinical relevance.

\subsection{The \textit{P}-value and Big Data: A Love-hate Relationship }

Bigger data provide a larger platform to make scientific enquiries, and, properly treated, may produce more consistent conclusions \citep{fan2014challenges, chen2019roles}. Between the loving moments, there are, sometimes, discordant days for the \textit{P}-value and big data.

First, big data may introduce big errors. Large-scale data such as magnetic resonance imaging (MRI) data may contain large-scale noise. For example, in fMRI data, multiples sources of noise, such as scanner-related noise including thermal noise and scanner instability noise, noise due to head motion and physiology, HRF model errors, and noise due to different sites, can corrupt the true signals \citep{chen2019roles}. There are three ways to mitigate this issue. First, one can aim to reduce noise before performing hypothesis testing. This calls for improving data acquisition and pre-processing. Second, scientists such as geneticists who consider a massive number of comparisons can improve reproducibility via cross-site and cross-study analyses (see meta-analysis below) and impose a very strict significance level (\textit{e.g.}, $P<5 \times 10^{-8}$) \citep{altshuler2005haplotype, pe2008estimation}. Third, even with extensive replication and strong signals, one may still observe false discoveries due to confounding variables or other biases. It is, therefore, important to improve general statistical thinking in practice (see \textbf{Section} 3.9).

Another problem with big data is the increasing likelihood to obtain spurious findings. Consider a hypothesis test to investigate the relationships between 500 brain edges and individual creativity scores. Among 500 edges under consideration, it is likely a few of them will be \textit{spuriously} associated with the outcome. This may introduce an erroneous scientific conclusion that these edges are underpinning creativity.

Thirdly, a small effect may appear significant, although not necessarily meaningful, with big data
\footnote{“In psychological and sociological investigations involving very large numbers of subjects, it is regularly found that almost all correlations or differences between means are statistically significant” \citep{meehl1967theory}.}. 
Empirically, a correlation of $0.1$ in a sample of $500$ has a \textit{P}-value around $0.025$; a correlation of $0.01$ in a sample of $100,000$ has a \textit{P}-value around $0.002$. The former is significant at $\alpha=0.05$, and the latter significant at $\alpha=0.005$, but the \textit{P}-values in these cases may suggest little. Indeed, in clinical trials and pathological studies, a small but significant effect size may offer little clinical inference and is difficult to interpret and reproduce \citep{miller2016multimodal}.

\subsection{The Misuses of the \textit{P}-value and Potential Remedies}

Here, we outline a few common misinterpretations and misuses of the \textit{P}-value and provide our modest recommendations. We do so in the form of questions and answers. Let us begin with a few questions:

\begin{enumerate} [(i)]
\item Should scientific conclusions be solely based on whether a \textit{P}-value is less than a specific threshold? Is \textit{post hoc} scientific interpretation based on the \textit{P}-value justified? 
\item	How could we prevent “\textit{P}-hacking”
\footnote{For example, conducting several statistical tests and only report those that pass the threshold, or add subjects to existing studies to lower the \textit{P}-value, in scientific discoveries.}?
\item	Many studies report results when observing a \textit{P}-value smaller than $0.05$ (or $0.005$). But is $0.05$ (or $0.005$) an optimal benchmark? 
\item	Does the \textit{P}-value measure the probability that the research hypothesis is true? Or does it measure the probability that observed data is due to chance? 
\item	Does obtaining a very small \textit{P}-value from hypothesis testing using a very large sample provide conclusive evidence? 
\item	Must scientific discovery always be accompanied by a hypothesis test and a \textit{P}-value? Are there alternative statistical approaches? 
\end{enumerate}

In \textbf{Table} 1, we attempt to answer these questions and present our modest recommendations. In \textbf{Section} 3.9 and \textbf{Fig}. \ref{Pipeline}, we present a flowchart to illustrate how to make better use of the \textit{P}-value in hypothesis testing.

\noindent 
\begin{sidewaystable*}
\label{Table_1}
\begin{tabularx}{\linewidth}{m{5cm} X }
\midrule
 \multicolumn{1}{ c }{Misuses and misconceptions of the \textit{P}-value}
& 
 \multicolumn{1}{ c }{Recommendations}\\
\midrule
        i. Scientific conclusions decisions are based only on whether a \textit{P}-value is less than a specific threshold.
     & 
        \parbox[c]{\hsize} {Observing that a \textit{P}-value is less than a threshold (\textit{e.g.}, $0.05$) alone does not, and should not, endorse a binary scientific conclusion. This is crucial when the \textit{P}-value is close to the threshold. For example, neither a rejection of a null hypothesis when \textit{P} $=0.045$ nor a failure to reject one when \textit{P} $=0.055$, offer conclusive evidence regarding the null, and need further including cross-validation, test-retest (\textit{e.g.}, permutation and bootstrap tests), and out-of-sample extrapolation (see \textbf{Figs}. \ref{Causal_inference} and \ref{Roles_of_the_P_value}). By further evidence, it means that if reporting a \textit{P}-value is mandatory (\textit{e.g.}, by a journal or a funding organization), reproducing a significant \textit{P}-value is highly recommended. For example, if a significant \textit{P}-value is discovered in a training sample, check if an independent testing sample also yields a significant \textit{P}-value. If modelling is concerned, verify whether fitted parameters obtained from a discovery sample can be extrapolated to a previously unseen testing sample. Extrapolation here means applying a trained model to new (testing) subjects to yield meaningful prediction \citep{woo2017building}, without further model fitting on the testing data.}\\
\midrule
    ii. “\textit{P}-hacking” (\textit{e.g.}, conducting several statistical tests, and only report those that pass the threshold).
     & 
      \parbox[c]{\hsize} {
      Instead of “hacking” the \textit{P}, (re)evaluate whether the experimental design is appropriate (\textit{e.g.}, is the design balanced? Is the sampling randomized? Is data collection appropriate? Is the data processing rigorous? Is the model suitable? Are all (statistical) assumptions met? If different experiments produce various types of data, conduct a meta-analysis and use the \textit{P}-value to combine evidence relating to a given hypothesis \citep{hedges1985statistical} (but see \textbf{Section} 5). Finally, if multiple statistical tests are conducted on the same data, report all analyses and their \textit{P}-values.
      } \\ 
\midrule
iii. 0.05 is the benchmark for the significance level.
&
\parbox[c]{\hsize} {
We cannot offer a strong recommendation for a benchmark of a significance level. The number 0.05 is coined by Fisher for convenience (see \textbf{Section} 2.2, and also see a recent call to use $0.005$ \citep{benjamin2018redefine, ioannidis2018proposal}). In general, when data is too small to be split into a training set and a test set, use a conservative significance level for confirmative discovery (\textit{e.g.}, $0.05$ is more conservative than $0.1$). Whenever possible, replicate the result in a novel sample. For large-scale data that can be split into a training set and a test set, consider a conservative significance level (\textit{e.g.}, $0.005$) for training, and a relatively more liberal one (\textit{e.g.}, $0.05$) for out-of-sample prediction.
}\\
\midrule
iv. \textit{P}-value measures the probability that the research hypothesis is true. \textit{P}-value measures the probability that observed data is due to chance.
&
\parbox[c]{\hsize} {
\textit{P}-value makes a statement about whether observed data supports a hypothetical research explanation. It does not give a statement about the explanation.
}\\
\midrule
v. (a) I have a very large sample.
\newline(b) I have conducted a hypothesis test and obtained a very small \textit{P}-value.
\newline(c) Thus, the result must be significant.
&
\parbox[c]{\hsize} {
\textit{P}-value is sensitive to sample size and variability in the sample. A very large sample size with a very small effect size can yield a significant \textit{P}-value. Such results may offer little inference in scientific studies and are likely to be irreproducible \citep{miller2016multimodal}. When facing large sample sizes, one could consider a data-driven approach instead (see \textit{vi}. below). If, however, a small but significant effect size is reproducible, the finding may still shed light on basic science, but it needs to be contextual (see \textit{i}. above). In biomedical studies, one could begin with a statistical statement, for example, “the difference was statistically significant”, followed by an additional statement on the clinical significance, using the effect size and their directions. 
}\\
\midrule
vi. Scientific discovery must be accompanied by hypothesis testing (and \textit{P}-value).
&
\parbox[c]{\hsize} {
There are alternative approaches. Depending on the specific scientific question, they are sometimes more suitable and feasible than hypothesis testing. On top of the \textit{P}-value, scientists can also report confidence, credibility, or prediction intervals to indicate effect size and direction. If scientists have prior knowledge about the problem, they could consider Bayesian methods. There are also alternative measurements for evidence, such as the \textit{likelihood ratio} or the \textit{Bayes Factor} (see \textbf{Section} 4). Finally, one could consider models based on decision theory and false discovery rates.
}\\
\midrule
\end{tabularx}
\caption{Common misinterpretation and misuse of the \textit{P}-values and recommendations \citep{benjamin2018redefine, gelman2014statistical, ioannidis2018proposal, nuzzo2014scientific, wasserstein2016asa, woo2017building}.}
\end{sidewaystable*}

\subsection{Making Better Use of the \textit{P}: An Improved \textit{P}(aradigm)}

Through our explorations, one may see the difficulty in suggesting a sample size, significance level, or power with which everyone agrees. A compromise can perhaps be done by suggesting a paradigm for conducting hypothesis testing aimed at improving reproducibility in scientific studies (see \textbf{Fig}. \ref{Pipeline} and a disclaimer therein). 

From \textbf{Fig}. \ref{Pipeline}, we see that even a seemingly simple associative analysis requires extra caution. We highlight that the interpretation of the \textit{P}-value is contextual. We need to interpret the \textit{P}-value along with, but never independent of, the research (experimental) design, hypothesis, the model and its assumptions, and prior evidence. Finally, it is important to improve statistical thinking and interdisciplinary training integrating statistical concepts and scientific explorations \citep{leek2015statistics}.

\section{Hypothesis Test in the Bayesian Realm}

Through Bayesian lens
\footnote{We added this section for completion purpose, as the discussion and debate between Bayesian evidence and \textit{P}-value persist. We, however, note that comparing two different types of evidence, namely Bayesian evidence and the \textit{P}-value, are like to compare two belief systems. Indeed, one colleague has summarized the discrepancy as follows:” I have always found the comparison between \textit{P}-values and posterior tail areas very puzzling because the \textit{P}-value is defined as a tail area where the value of the sample changes, but the posterior tail area varies over different values of theta (the unknown parameter). How are these two comparable unless we want to compare eggs with sausages?”}
, the posterior probability of $H_0$ given data $x$ provides an alternative way to gather evidence \citep{bayarri2000p, berger1997unified, berger1987testing, casella2021statistical, diamond1983clinical, dickey1977tail, held2018p, shafer1982lindley}.

The major commonality between the \textit{P}-value and Bayesian evidence is that both are defined \textit{post-data}. Both, therefore, are data- or sample-specific and provide evidence as much as the data suggest. When a uniform prior (perhaps the most extreme case of a non-informative prior) and a Gaussian likelihood are employed, it is relatively easy to see that the \textit{P}-value and the Bayesian provide the same information. There are, however, two major differences between them. First, Bayesian evidence incorporates prior information into its model whereas the \textit{P}-value does not. Second, the \textit{P}-value is calculated where the parameter is an unknown but fixed value; Bayesian evidence is calculated where the parameter is a random variable. 

Unlike the \textit{P}-value, which is largely determined by the observations and the statistical model $M_\theta (\bx)$, Bayesian evidence depends not only on the observations and the model, but also on the prior. In other words, if one has a strong prior (for example a very large precision compared to it of the likelihood), no matter how much information the data contain, the posterior parameters are chiefly dictated by the prior. The hypothesis testing outcomes are, consequently, chiefly determined by the density function of the prior. On the other hand, weak prior surrenders to data; the posterior parameters are, therefore, closer to the MLE of the likelihood. Consequently, the hypothesis testing outcomes may be chiefly determined by the likelihood function. To see this more concretely, let's consider an example. 

Suppose we have some prior knowledge about a parameter $\mu \sim N(\mu_\pi, \sigma^2_\pi)$. The likelihood of drawing data $\bx = (x_1, x_2, \ldots, x_n)$ is $\mathbb{P}(\bx \vert \mu_x) = \prod_{i=1}^n \mathbb{P}(x_i \vert \mu_x)=(2 \pi \sigma^2_x)^{-n/2} \exp \bigg\{$  $-\dfrac{1}{2\sigma_x^2}$ $\sum_{i=1}^n (x_i - \mu_x)^2 \bigg \} $. It follows that after seeing data $\bx$, the posterior distribution of $\mu$ becomes $\mu \vert \bx \sim N(\mu_n, \sigma_n^2)$, where $\mu_n = \left( \dfrac{n}{\sigma_x^2} + \dfrac{1}{\sigma_\pi^2} \right )^{-1} $ $\left [\dfrac{n}{\sigma_x^2} \left (\dfrac{\sum_{i=1}^n x_i}{n} \right) + \dfrac{1}{\sigma_\pi^2} \mu_\pi
\right]$, and $\sigma_n^2 = \left (
\dfrac{n}{\sigma_x^2} + \dfrac{1}{\sigma_\pi^2}
\right )^{-1}$. Suppose $\mu_\pi = 90$ and $\mu_x = 110$, and we wish to examine two sets of hypotheses:
\begin{eqnarray}
(S1): \hspace{2mm} H_0: \mu \leq 109 \hspace{3mm}  \textit{vs.}  \hspace{3mm}H_1: \mu > 109; \nonumber\\
(S2): \hspace{2mm} H_0: \mu \leq 111   \hspace{3mm} \textit{vs.} \hspace{3mm} H_1: \mu > 111.\nonumber
\end{eqnarray} 

Let's consider scenarios that cover three basic relationships between the precision of the prior and it of the likelihood: 
\begin{enumerate}
\item The prior is more precise (with a smaller standard division) than the likelihood, \textit{i.e.}, $\sigma_\pi:\sigma_x = 1:5$; 
\item The prior has similar precision as the likelihood, \textit{i.e.}, $\sigma_\pi:\sigma_x = 5:5$; 
\item The prior is less precise than the likelihood, \textit{i.e.}, $\sigma_\pi:\sigma_x = 5:1$. 
\end{enumerate}

\begin{figure*}[h!]
\makebox[\textwidth][c]{\includegraphics[width=1\textwidth]{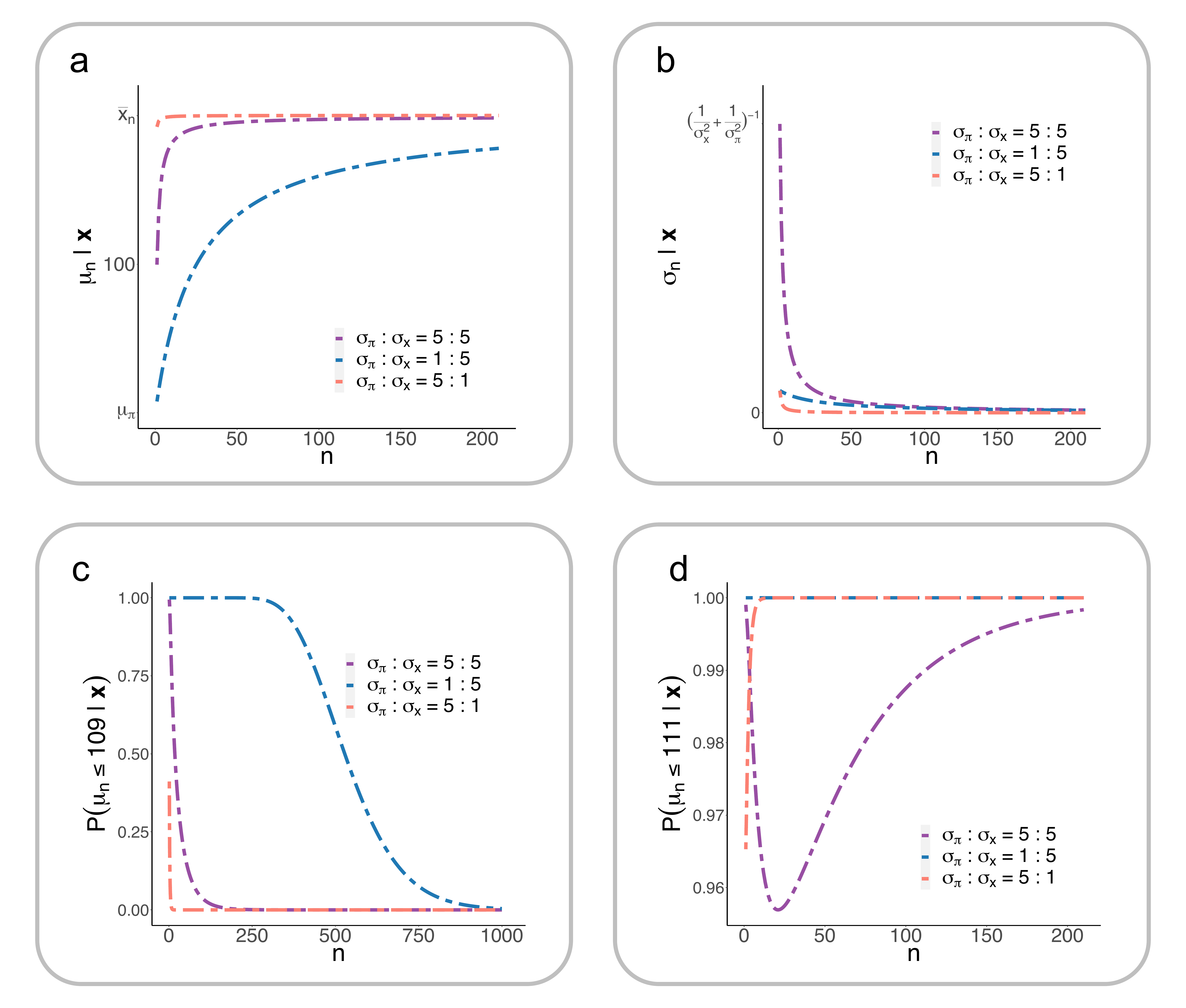}}
\caption{The behaviour of Bayesian posterior evidence.}
\medskip
\parbox[c]{\hsize} {(a) The behaviour of the posterior mean. (b) The behaviour of the posterior standard deviation. (c) The behaviour of posterior evidence for $H_0: \mu \leq 109$. (d) The behaviour of posterior evidence for $H_0: \mu \leq 111$. See text for explanations. 
}
\label{Bayesian_evidence}
\end{figure*}

In all scenarios (see \textbf{Fig}. \ref{Bayesian_evidence} (a)-(b)), the posterior mean approaches to the mean of the likelihood function (which equals to the MLE) as more data are gathered (\textit{i.e.}, as $n$ increases), and the variance (which determines our confidence about the accuracy of the posterior estimate) decreases towards zero. The rate of convergence (loosely speaking), however, differs across the three scenarios. When the prior is less precise than the likelihood (\textit{e.g.}, $\sigma_\pi: \sigma_x=5:1$), the posterior mean converges to $\bar{x}_n$ (the MLE) rather quickly with smaller a posterior variance, as the dominating information is provided by the data. When the prior is more precise than the likelihood (\textit{e.g.}, $\sigma_\pi: \sigma_x=1:5$), the posterior mean is quite resistant to converging to $\bar{x}_n$ (the MLE) with a larger a posterior variance; unless a lot more data are used, the posterior takes into significant consideration of the prior. The third case, where prior and the likelihood have similar convergence rate and variance are both moderate compared to the other two cases.  

The usefulness of Bayesian evidence lies in that it balances (or modulates) prior knowledge and knowledge gathered from the data (the likelihood). Let's consider the first set of hypotheses (S1): $H_0: \mu \leq  109$   \textit{vs.}   $H_1: \mu >109$. The mean of the prior ($\mu_\pi=90$) is in favour of the null but the likelihood is not ($\mu_x=110$). When the prior is not as precise as the likelihood (\textit{e.g.}, $\sigma_\pi: \sigma_x=5:1$), the posterior mean gives up supporting for the null immediately after seeing the data from a distribution centred at 110 (see the red line in \textbf{Fig}. \ref{Bayesian_evidence} (c)). When prior is more precise than the likelihood (\textit{e.g.}, $\sigma_\pi: \sigma_x=1:5$), however, it requires more data to convince the posterior mean (see the blue line in \textbf{Fig}. \ref{Bayesian_evidence} (c)).  

Next, let us consider the second set of hypotheses (S2): $H_0: \mu \leq  111$   \textit{vs.}   $H_1: \mu >111$. Both means of the prior ($\mu_\pi=90$) and the likelihood ($\mu_x=110$) are in favour of the null, and the prior is more in favour of the null than the likelihood. When prior is more precise than the likelihood (\textit{e.g.}, $\sigma_\pi: \sigma_x=1:5$), the posterior went to support the null right after seeing a small number of data points (see the blue line in \textbf{Fig}. \ref{Bayesian_evidence} (c)). When the prior is not as precise as the likelihood (\textit{e.g.}, $\sigma_\pi: \sigma_x=5:1$), the posterior mean needed more data to support the null (see the red line in \textbf{Fig}. \ref{Bayesian_evidence} (c)). The interesting scenario is when both the prior and the likelihood are not precise ($\sigma_\pi: \sigma_x=5:5$), in this case, it takes some battling, after seeing more data, to achieve consensus regarding supporting the null (see the purple line in \textbf{Fig}. \ref{Bayesian_evidence} (d)).

\subsection{On Bayesian Posterior Evidence \textit{vs.} the \textit{P}-value}

\begin{quote}
[M]ost nonspecialists interpret $[\textit{P}]$ precisely as $\mathbb{P}(H_0 \vert x)$ (see \citep{diamond1983clinical}) [thereby committing the fallacy of the transposed conditional, our insertion], which only compounds the problem \citep{berger1987testing}.
\end{quote}

As argued in \textbf{Section} 2.3, one chief difference between the \textit{P}-value and Bayesian evidence is that the former considers the parameter as an unknown but fixed value and the latter considers the parameter as a random variable on which a probability distribution can be imposed. Naturally, it is difficult to compare an argument built on a fixed number with one built on a distribution. Yet, as Bayesian evidence leverages between information provided from the prior and it from the data when the prior brings in little information (\textit{i.e.}, non-informative), it seems possible that the Bayesian evidence may deliver the same amount of information - from the data alone - as the \textit{P}-value does. 

To see the last point above more vividly, let's consider an extreme case. Consider a uniform (improper) prior defined on the real line and Gaussian likelihood $\bx \vert \mu_x \sim N(\mu_x, \sigma_x^2)$. The posterior then is $ \mu \vert \bx \sim N(\mu_b, \sigma_n^2 )$, where $\mu_n= \bar{x}_n$, and $\sigma_n^2=(\sigma_x^2)/n$. For a null hypothesis $H_0: \mu \leq \mu_0$, the Bayesian evidence is $\mathbb{P}( \mu \leq \mu_0 \vert x) = \mathbb{P} \left(
\dfrac{\mu - \bar{x}_n}{\sigma_x / \sqrt{n} } 
\leq 
\dfrac{\mu_0 - \bar{x}_n}{\sigma_x / \sqrt{n} } 
  \bigg \vert \bx \right ) = \Phi \left(\dfrac{\sqrt{n} (\mu_0-\bar{x}_n) }{\sigma_x}
\right ) $, where $\Phi$ is the CDF for $N(0,1)$. For the same hypothesis, the \textit{P}-value is $1 - \mathbb{P}(\Delta(\bX) > \Delta(\bx), \mu = \mu_0) = \mathbb{P} \left (
\Delta(\bX) \leq \dfrac{ \sqrt{n}(\mu_0-\bar{x}_n)}{\sigma_x}
\right ) =\Phi \left (
\dfrac{ \sqrt{n}(\mu_0-\bar{x}_n)}{\sigma_x}
\right)$.

Naturally, one would ask which is more suitable in scientific studies? A definitive answer is difficult. In general, one should consider the Bayesian approach if one has some strong \textit{a priori} belief about the parameter and considers the \textit{P}-value when one only has data without prior knowledge about them. There are, however, a few general consensuses regarding the amount of evidence (not comparing the two systems, our emphasis) they provide. Since one- and two-sided hypothesis tests (but see \textbf{Remark} 3 in \textbf{Section} 2) are the predominant practices in scientific and clinic expositions, we will focus on these two types of tests in the following. Readers who are interested in composite hypothesis tests could refer to \citep{bayarri2000p, berger1997unified}.

\begin{itemize}
\item[] 1. [For a two-sided (point null) test]: The \textit{P}-value tends to overstate the evidence against the null \citep{dickey1977tail, shafer1982lindley, berger1987testing}; that is, the \textit{P}-value is smaller than the Bayesian posterior evidence.
\item[] 2.a [For a one-sided test]: The \textit{P}-value can be approximately equal to the Bayesian posterior evidence \citep{pratt1965bayesian}.
\item[] 2.b [For a one-sided test]: One can construct a (improper) prior such that the \textit{P}-value and the Bayesian posterior evidence match \citep{degroot1973doing}.
\item[] 3.a [For a one-sided test]: For data following distribution with a monotone likelihood ratio, and that has unimodal density, symmetric about zero, or is normal $(0, \sigma^2 )$, where $0<\sigma^2< \infty$, the \textit{P}-value is equal to $ \inf \mathbb{P} (H_0  \vert x)$, where the infimum is taken over a class of priors \citep{berger1987testing}. 
\item[] 3.b [For a one-sided test]: For other distributions, the \textit{P}-value is greater than or equal to $ \inf \mathbb{P} (H_0  \vert x)$, suggesting that the \textit{P}-value may be understating the evidence against the null \citep{berger1987testing}.
\item[] 4. If a prior mass is concentrated at a point (or in a small interval) and the remainder is allowed to vary over the alternative hypothesis $H_1$ (in other words one has strong prior information), then there could be a (noticeable) discrepancy between the Bayesian posterior evidence and the \textit{P}-value. To see this, suppose there is some prior information about the location parameter $\theta: \pi(\theta) = \pi_0 h(\theta)+(1-\pi_0 )g(\theta)$. This is equivalent to putting a prior $\pi_0$ to $\theta=0$ and another $(1-\pi_0 )g(\theta)$ to $\theta>0$, assigning mass on the point null hypothesis, thereby biasing the prior in favour of $H_0$ (for any fixed $g$) \citep{berger1987testing}.
\end{itemize}

Taken together:

\begin{enumerate}[\hspace{5mm}(i)]
\item For a two-sided test (\textit{e.g.}, testing whether the disease prevalence is above or below 20\%), the conclusions made using Bayesian evidence may be more conservative than using the \textit{P}-value \citep{dickey1977tail, shafer1982lindley, berger1987testing}. 
\item For a one-sided test (\textit{e.g.}, testing whether the disease prevalence is above 20\%), the two offer approximately the same evidence (and can be constructed to be equivalent) \citep{pratt1965bayesian, degroot1973doing}.
\item When one has strong prior information (say about the null hypothesis), the Bayesian alternatives would favour the null \citep{berger1987testing}. It is particularly attractive, for example, when one conducts region fine-mapping to identify the true causal variant(s) \citep{stephens2009bayesian}. 
\item When samples are large, the small \textit{P}-values (see discussions above and \textbf{Fig}. \ref{Lady_tasting_tea}) almost systematically reject the null; the Bayesian alternatives do not \citep{kass1995bayes}. 
\item One should be cautiously aware that if different studies adopt different priors, it would be problematic to compare findings between studies \citep{fadista2016famous}.
\end{enumerate}

\subsection{An Example: The \textit{Bayes Factor} in Model Comparison}

It is not always necessary to report the Bayesian evidence in the context of posterior probability; rather, one can report the ratio of the posteriors of two hypotheses. A useful application is to perform a model comparison. Suppose an epidemiologist is interested in investigating whether the incidence rate of a disease is at 20\% ($H_1$), or at 10\% ($H_2$).

More concretely, suppose $H_1$ and $H_2$ are two hypothesized models parameterized by $\theta_1$ and $\theta_2$, respectively. The \textit{Bayes factor} (see \citep{kass1995bayes} for a comprehensive discussion), or $K$, is written as:
\begin{eqnarray}
\label{eqn:3}
K &=& \dfrac{P(x \vert H_1)}{P(x \vert H_2)}
= \dfrac{\int P(\theta_1 \vert H_1) P(x \vert \theta_1, H_1) d \theta_1}{\int P(\theta_2 \vert H_2) P(x \vert \theta_2, H_2) d \theta_2}   
\\
    & = & \dfrac{P(H_1 \vert x)}{P(H_2 \vert x)} \times \dfrac{P(H_2)}{P(H_1)} \nonumber
\end{eqnarray}
where $x$ stands for the data, and $H_1$ and $H_2$ are two hypothesized models. Note that when the priors $P(H_1)$ and $P(H_2)$ are equal, the \textit{Bayes factor} reduces to $K=\dfrac{P(H_1 \vert x)}{P(H_2 \vert x)}$), thus degenerating to a \textit{likelihood ratio} test.

In words, the \textit{Bayes factor} compares how likely the data are generated from model 1 ($H_1$) as compared to model 2 ($H_2$); hence the larger the $K$, the stronger evidence the data support $H_1$ over $H_2$. To see it more concretely, suppose the epidemiologist wanted to test the prevalence of a certain type of disease. The epidemiologist came up with six candidate models ($H_1$) to test against an alternative model which assumed the prevalence was at 10\% ($H_2$, where, for example, one considered a parameter $\theta_2=0.10$). The six candidate models considered their parameters as follows: (1) a uniform distribution or $\theta_1  \sim U[0,1]$; (2) 15\%, or $\theta_1=0.15$; (3) 40\%, or $\theta_1=0.40$; (4) 20\%, or $\theta_1=0.20$ (which is the maximum likelihood estimator (MLE)); (5) a Normal distribution or $\theta_1  \sim  N(0.2, 0.01)$, which can be considered as the MLE plus a small noise; (6) a Normal distribution or $\theta_1  \sim N(0.2,0.1)$, which can be considered as the MLE contaminated by a large noise, say, due to sampling error. 

\begin{figure}[h!]
\makebox[\textwidth][c]{\includegraphics[width=1.2\textwidth]{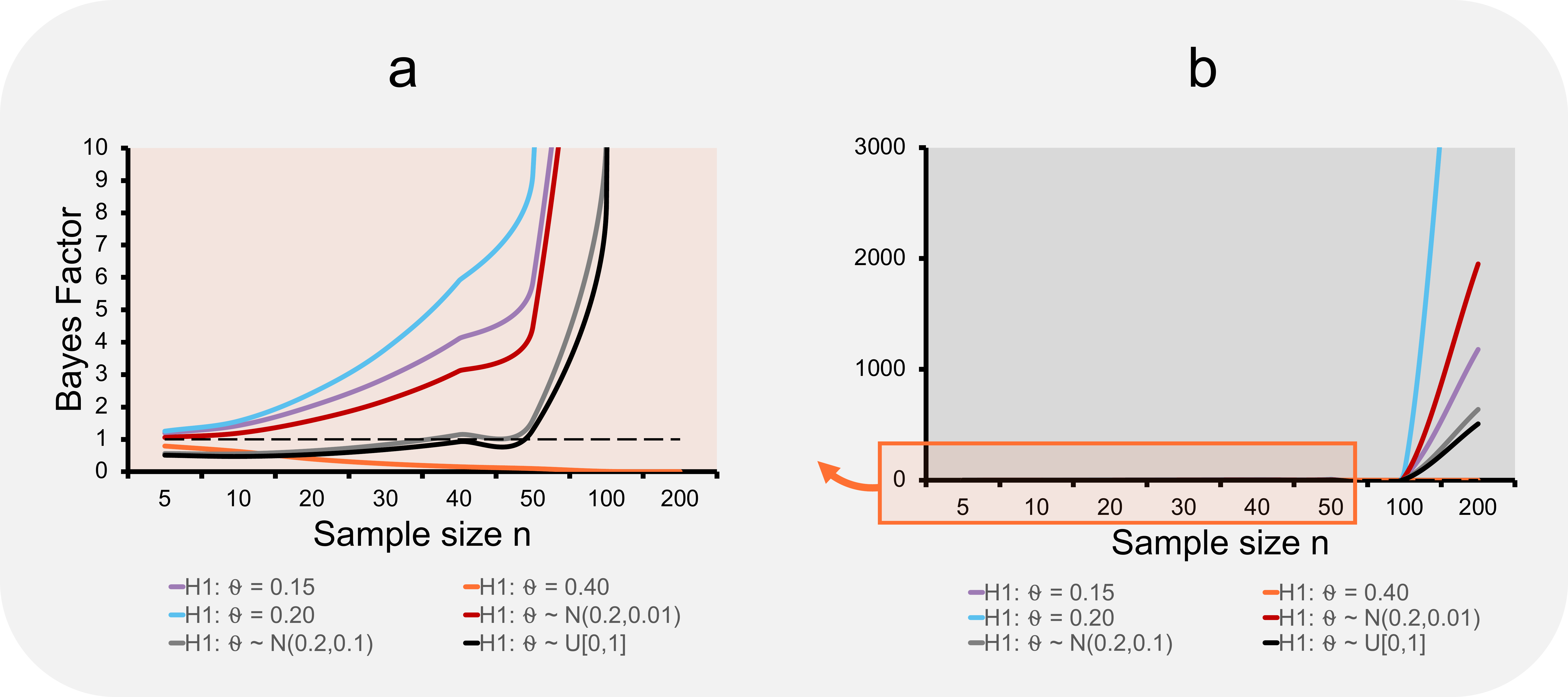}}
\caption{An illustration of the \textit{Bayes factor} in model comparison.}
\medskip
\parbox[c]{\hsize} {Consider an experiment comparing two models $H_1$ and $H_2$. For simplicity, the sample incidence rate was fixed at $0.2$, no matter of sample size (that is, 8 for a sample of 40, and 20 for a sample of 100). Figure (a) is the zoomed-in snapshot of the orange box in Figure (b). \textbf{Figure (a)} shows how the \textit{Bayes factor} changes when the sample size was smaller than $100$. \textbf{Figure (b)} shows how the \textit{Bayes factor} behaves when the sample size was larger than $100$. The experiment considered six candidate models for $H_1$ with the prevalence parameterized as, respectively, (1) from a uniform distribution or $\theta_1  \sim U[0,1]$; (2) 15\%, or $\theta_1=0.15$; (3) 40\%, or $\theta_1=0.40$; (4) 20\%, or $\theta_1=0.20$ (which is the maximum likelihood estimator (MLE)); (5) from a Normal distribution or $\theta_1  ~ N(0.2,0.01)$, which can be considered as the MLE plus a small noise; (6) from a Normal distribution or $\theta_(1 )\sim N(0.2,0.1)$, which can be considered as the MLE contaminated by a large noise, say, due to sampling error. The alternative model $H_2$ had a parameter $\theta_2=0.10$. The $H_1$ whose hypothesized parameter equalled the sample incident yielded the largest \textit{Bayes factor}. In other words, the maximum likelihood estimator or MLE (in this case $0.2$) achieved the optimal \textit{Bayes factor}. The results also showed that the farther a hypothesized parameter departed from the MLE (\textit{e.g.}, $\theta_1=0.40$ is farther from $0.2$ than $\theta_1=0.15$), the smaller the \textit{Bayes factor} (or evidence); this was true no matter of sample size; but the larger the sample size, the stronger the evidence. When the sample size was small, the model with $\theta_1  \sim N(0.2,0.01)$ (namely the MLE plus some Gaussian noise $N(0.2,0.01)$ underperformed $\theta_1=0.15$, indicating the noise had contaminated the evidence. With larger sample size, the former outperformed the latter, indicating the signals from large-scale data had overcome the noise. 
}
\label{Bayesian_model_comparison}
\end{figure}

The epidemiologist considered several samples of sizes 5, 10, 20, 30, 40, 50, 100, and 200. For comparison, suppose that the true incident rates were all at 20\%; namely for each sample, there were, respectively, 1, 2, 4, 6, 8, 10, 20, and 40 patients. Using Equation \eqref{eqn:3}, the \textit{Bayes factor} for each test is calculated and presented in \textbf{Fig}. \ref{Bayesian_model_comparison}.

There are four messages we can draw from the simulation studies, from which one could peer into the general behaviour of the \textit{Bayes factor}. 
\begin{enumerate}[\hspace{5mm}(a)]
\item When the hypothesis (in $H_1$) is close to the truth (20\%), the \textit{Bayes factor} uniformly supports $H_1$ over $H_2$ (as the \textit{Bayes factor} is larger than 1 no matter the sample size). 
\item When the hypothesis (in $H_1$) is far from the truth, the \textit{Bayes factor} uniformly opposes $H_1$ over $H_2$ (as the \textit{Bayes factor} is no larger than 1 no matter the sample size). 
\item The larger the sample size, the stronger evidence the \textit{Bayes factor} provides for supporting (or opposing) $H_1$. This is a major difference from the \textit{P}-value, which uniformly decreases when the sample size increases. 
\item The \textit{Bayes factor} accounts for prior information and uncertainties in the model. For example, when prior information about $\theta_1$ is close to the truth (20\%), the \textit{Bayes factor} strongly supports $H_1$; when the prior is contaminated by some noise (as in $N(0.2,0.01)$ and $N(0.2,0.1))$, the \textit{Bayes factor} becomes smaller, and the more noise found in the prior the smaller the \textit{Bayes factor}. When there is uncertainty (as in a uniform distribution), the small sample size would support $H_2$ (namely $\theta_2=0.1$); when sample size becomes sufficiently large, the \textit{Bayes factor} detects from the data that it is increasingly unlikely that the data correspond to a model ($H_2$) where $\theta_2=0.1$.
\end{enumerate}

\section{Pooling \textit{P}-values via Meta-analysis}
The analysis of large-scale datasets in scientific studies has, in general, two advantages: information accumulation and commonality extraction \citep{chen2019roles}.

Information accumulation includes increasing the size of a single dataset and combining different datasets. For the former, it expands the sample size (by collecting more subjects), temporal dynamic (by obtaining more longitudinal measurements for each subject), and spatial variability (by increasing the number of features collected or areas measured for each subject). For the latter, it compounds heterogeneous samples, disease categories, or task paradigms. \textit{P}-values obtained from large-scale datasets may more clearly suggest the difference between subpopulations (\textit{e.g.}, healthy versus disease, male versus female, and individuals under various treatments or stimuli versus controls), and identify the pathological-, gender-, treatment-, and task-specific phenotypes. Hypothesis testing and the \textit{P}-value obtained from repeated measurements help to delineate the longitudinal changes of the features, thereby potentially improving disease assessment over time, and paving the way for longitudinal disease prediction and progression monitoring \citep{ramsay1997functional, giedd1999brain, casey2000structural, johnson2001functional}.

Commonality extraction refers to obtaining converging evidence from multiple studies and datasets. On the one hand, data sets obtained from different studies and experimental conditions contain heterogeneous signals. On the other hand, they may be subject to different degrees of systematic bias due to different experimental designs (\textit{e.g.}, a complete factorial design versus a fractional factorial design \citep{wu2011experiments}), noises (such as head motion \citep{ciric2017benchmarking}), measurement errors due to data aggregation under different paradigms and from different sites \citep{cao2019toward}, missing data \citep{little2019statistical}, and reporting bias (for example, only positive results are reported or published \citep{ioannidis2014publication}). Consequently, data analysis results reported from mis-specified models \citep{mayo2006severe} or datasets obtained under different designs and conditions may provide different \textit{P}-values, thereby generating different, sometimes opposite conclusions (also see the Simpson's paradox). 

Today, it is increasingly common to see studies considering and balancing both information accumulation and commonality extraction. For example, a committee of researchers may organize several study groups conducting multiple experiments and gathering data at different locations under various conditions, a good practice that has already been adopted in clinical trials (multicentre studies), to seek for converging evidence that may address a common scientific question. Naturally, one would ask, is there a suitable approach to obtain evidence from aggregated studies and datasets?

\begin{figure}[h!]
\makebox[\textwidth][c]{\includegraphics[width=1.4\textwidth]{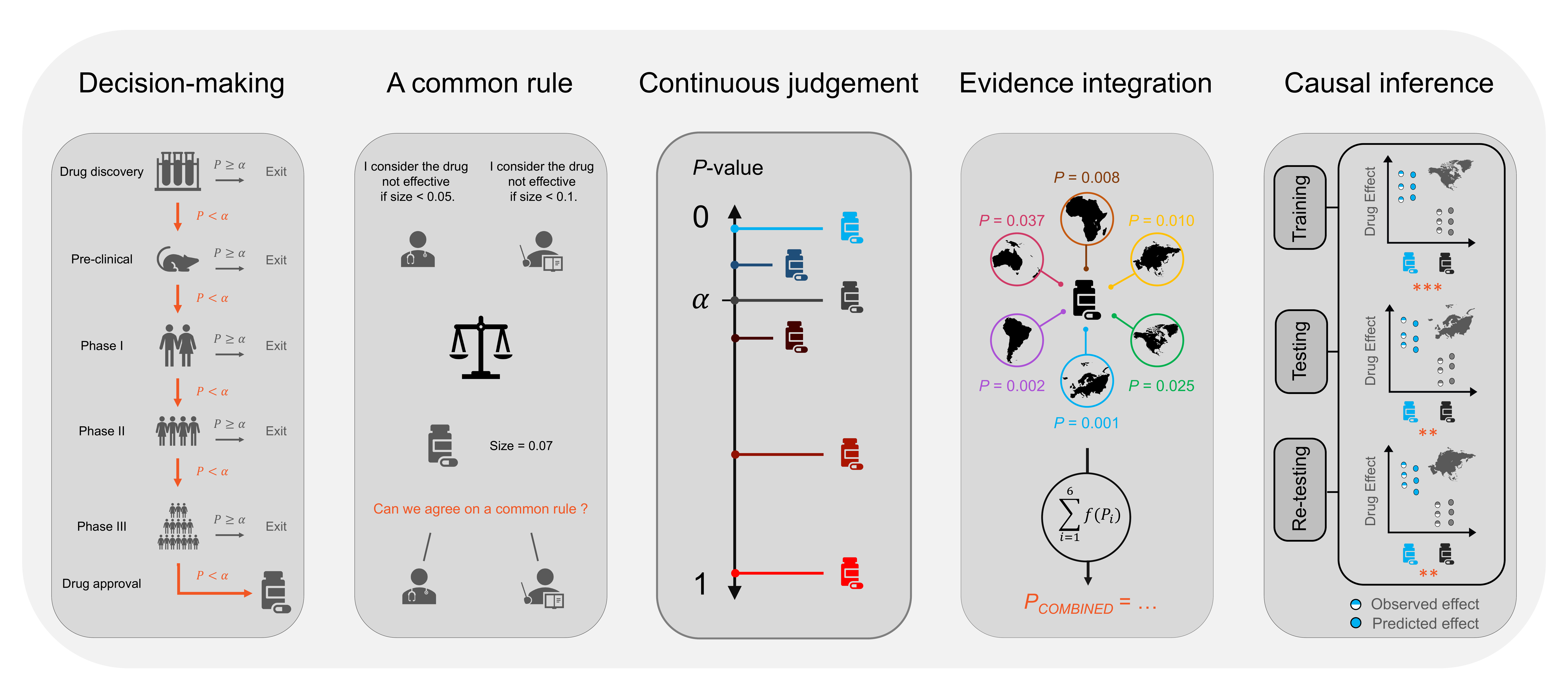}}
\caption{Key useful roles of the \textit{P}-value.}
\parbox[c]{\hsize} {From left to right: (a) The \textit{P}-value helps to form a simple, clear, and perhaps universally agreeable decision-making system. It has been accepted by a broad scientific, clinic, and medical communities. (b) It provides a common, and straightforward rule that guides multiple experimenters to evaluate and compare findings based on respective \textit{P}-values and a pre-agreed significance level. (c) It evaluates the outcomes of a test on a continuous scale. (d) It allows for, although with caution (see \textbf{Section} 5), integrating results from multiple studies and datasets. (e) It facilitates causal enquires and provides a metric to evaluate and determine the existence and strength of potential causation (see \textbf{Section} 3.3 and \textbf{Fig}. \ref{Causal_inference} for more details).  
}
\label{Roles_of_the_P_value}
\end{figure}

The \textbf{meta-analysis}
\footnote{The term meta-analysis (analysis of analyses) was coined by Gene V. Glass in 1976 in \textit{Primary, secondary, and meta-analysis of research}.}
(analysis of analyses) is a useful approach to integrate and extract evidence from large-scale heterogeneous datasets, reduce reporting bias, and draw potentially reliable conclusions. There are, however, both advantages and risks of pooling \textit{P}-values from different studies and datasets. They can be generally present in four areas.

\begin{enumerate} [\hspace{5mm}(1)]
\item Meta-analysis can integrate results from different studies. For example, Fisher's combined probability test integrates the \textit{P}-values obtained from multiple studies and datasets (see \textbf{Fig}. \ref{Roles_of_the_P_value}).
\item Meta-analysis may reduce bias. For example, when regions of (prior) interest have more liberally thresholds than others (such as in large-scale neuroimaging studies), the results are likely biased towards these regions. \textit{Seed-based d mapping} (also known as the \textit{signed differential mapping} (SDM)) \citep{radua2009voxel, radua2012new} can (meta)analyse functional and structural brain data across multiple large-scale (neuroimaging) studies
\footnote{First, peak coordinates (\textit{e.g.}, the brain regions where the differences between healthy and disease are the highest) are combined with \textit{t}-statistic maps (each \textit{t}-statistic map can be plotted to the brain space where regions with large t values indicate activation) from studies using SPM; second statistical maps and effect-sizes maps are recreated; finally, individual maps are combined according to intra-study variance (\textit{i.e.}, studies with large sample sizes and/or lower error contribute more) and inter-study heterogeneity (\textit{i.e.}, studies with large variances contribute less).}
to reduce bias and improve power. 
\item Meta-analysis can examine whether discoveries are reproducible. Meta-analysis can perform a leave-one-study-out cross-validation in the spirit of a leave-one-subject-out cross-validation. For example, it first compares the estimate (\textit{e.g.}, mean activation of a brain lesion) from one study to the summarized estimate from the remaining ($n-1$, where n is the number of total studies) studies, and then iterates the process and judges, via the \textit{P}-value, whether the conclusion made across the studies are reliable and reproducible
\footnote{Note that point estimates are not reproducible, even under ideal conditions - they represent a single value from a range of possible values represented by the relevant sampling distribution. Hence, the claim that for an optimal estimator $t(X)$, the estimate $t(x_0)$ approximates the true value of $\theta$ is unwarranted \citep{spanos2019probability}.}.
\item Meta-analysis may be \textit{inappropriate} in practice unless all \textit{P}-values share (approximately) the same statistical context (including the model, framing of hypotheses, sample size, \textit{etc}.), whose shared statistical model is adequate. Pooling \textit{P}-values from statistically mis-specified models can be dangerous \citep{spanos2015error}.
\end{enumerate}

\section{Conclusion}
In this paper, we aimed to discuss the roles and challenges of the \textit{P}-value in hypothesis testing. We first outlined the roles the \textit{P}-value plays in scientific studies, and discussed the associations between the \textit{P}-value, sample size, significance level, and statistical power. Subsequently, we presented common misuses and misinterpretations of the \textit{P}-value, accompanied by modest recommendations. To complement our discussion, we compared statistical significance and clinical relevance. Additionally, we presented the Bayesian alternatives of seeking evidence. Finally, we discussed the potential usefulness and risks of performing meta-analysis to integrate and extract evidence from multiple studies and datasets. 

To summarize, hypothesis testing and the \textit{P}-value form a decision-making system; they provide a common, simple rule that guides experimenters, evaluating and comparing findings via the \textit{P}-values; they help to examine test outcomes on a continuous scale; they enable, with caution, integrating results from multiple studies and datasets; they facilitate causal enquires and provide a metric to evaluate and determine the existence and strength of potential causation. Today, they are supporting scientific enquires to test the relationship between group-specific, idiosyncratic, genetic, and environmental features, the difference between outcomes from multiple geographical (such as corps from different fields) and biological (such as patterns from different brain areas), how external stimuli and environmental factors affect genetic organizations and biological characteristics (such as heart rates and brain signals), how these patterns underpin human behavior, and how their irregularity may lead to malfunction and illnesses.

We believe that the \textit{P}-value will continue to play important roles in hypothesis-testing-based scientific enquires, whether in its current form or modified formulations. We also believe that there will be a continued effort to seek more rational ways to extract knowledge from data and more holistic interpretation for statistical and scientific evidence.

As the employment of hypothesis testing and \textit{P}-values is and will for the foreseeable future remain one of the standard practices in scientific enquiries, a beginning can perhaps be made by improving our understanding of its roles, weaknesses, and misuses. Our discussions highlight that its applications and interpretation must be contextual, considering the scientific question, experimental design (including the model specification, sample size, and significance level), statistical power, effect size, and reproducibility of the findings. We are rewarded if our explorations have brought you some insights to your current and future studies.

\begin{acks}[Acknowledgments]
O.Y.C. wrote the paper with comments from all other authors. J.D. wrote Section 4.2. The authors thank Aris Spanos for his constructive comments regarding two early versions of the paper. 
\end{acks}

\begin{funding}
Non-declared.
\end{funding}

\bibliographystyle{imsart-nameyear.bst}
\bibliography{reference.bib}       

%
%
%
%
%
%
%
%
%
%

\end{document}